\documentclass{article}
\usepackage{arxiv}

\usepackage[utf8]{inputenc} 
\usepackage[T1]{fontenc}    
\usepackage{hyperref}       
\usepackage{url}            
\usepackage{booktabs}       
\usepackage{amsfonts}       
\usepackage{nicefrac}       
\usepackage{microtype}      
\usepackage{lipsum}
\usepackage{bm}
\usepackage{graphicx}
\usepackage{enumerate}
\usepackage{siunitx}
\usepackage[]{algorithm2e}
\usepackage{amsmath}
\usepackage{amsbsy}
\usepackage{amssymb}
\usepackage{color}
\usepackage{subfigure}
\usepackage{wrapfig}
\usepackage{appendix}
\usepackage{tikz}
\newcommand*\circled[1]{\tikz[baseline=(char.base)]{
		\node[shape=circle,draw,inner sep=2pt] (char) {#1};}}
\title{Centre mode instability of a dilute particle-laden swirling jet in a swirl flow combustor}

\author{
  Srikumar Warrier \thanks{Corresponding author.}\\
  Interdisciplinary Centre for Energy Research\\
  Indian Institute of Science\\
  Bangalore 560012, India. \\
  \texttt{srikumarw@alum.iisc.ac.in} \\
\And
Gaurav Tomar  \\
Department of Mechanical Engineering\\
Indian Institute of Science\\
Bangalore 560012, India. \\
}

\begin{document}
\maketitle

\begin{abstract}
Linear stability of a locally parallel annular swirling jet laden with particles in a swirl flow combustor is considered.  At low Stokes numbers, the eigenspectra of the particle-laden jet with uniform particle concentration shows three unstable modes namely centre, sinuous and varicose modes. As the Stokes number is increased to unity, the growth rates of the centre and shear layer modes reduces compared to that of the unladen swirling jet.  The magnitude of the velocity eigenmodes peaks in the vortex core and decays radially outward. The variation in particle concentration occurs mostly in the vortex core and almost none in the shear layer. The strength of flow reversal at the jet centreline is given by the backflow parameter. An increase in the backflow parameter increases the growth rate of the centre mode. Non-uniformity in the base-state particle concentration is introduced using a Gaussian function varying in the radial direction and a reduction in the growth rate of the centre mode is seen compared to the uniform particle concentration profile.  When the location of the peak of the base-state particle concentration profile is inside the vortex core, the centre modes are stable. Linearized vorticity budget analysis reveals that this is accompanied by a decrease in the net generation of perturbation vorticity in the axial direction and increased radial and azimuthal perturbation vorticity. 
\end{abstract}

\keywords{ Swirling jet \and linear stability analysis \and
centre modes, sinuous modes, varicose modes \and ring modes  }

\section{Introduction}
\label{sec:Intro}
Round jets and swirling jets find application in aerospace propulsion devices such as gas turbine combustors. In most of the combustors used now-a-days, swirling jets have been employed for flame stabilization, which is essential to prevent flame blow-off. At certain conditions (large swirl numbers), swirling jets can generate flow reversal in the near field of the nozzle, thereby aiding in flame stabilization. These are called swirl-stabilized combustors. Additionally, they also aid in mixing of fuel and air in the combustor. In the context of design of propulsion devices for aerospace applications, the stability of open shear flows like planar jets, round jets and swirling jets of gas/air carrying atomized fuel droplets (considered as small particles) plays a considerable role in determining the efficiency of combustion. Eaton and Fessler\cite{EATON_1994} showed that particle-laden turbulent shear layers exhibit preferential concentration of particles. 
They further show that preferential concentration has a significant effect on the turbulence of the carrier fluid. Experimentally, it has been observed that preferential concentration of particles is mainly caused by coherent structures in turbulent flows where dense particles are centrifuged away from the vortex cores and accumulate in the vicinity of the shear layers. For instance, in dilute particle-laden mixing layers, particles accumulate in the braid region, outside the vortex cores, as observed by Meiburg et al.\cite{meiburg_2000}. DeSpirito et al.\cite{DESPIRITO_2001} performed linear stability analysis of a particle-round jet subjected to axi-symmetric perturbation ($m=0$), and concluded that at small Stokes numbers, the unladen jet is more unstable than the particle-laden jet. At intermediate Stokes numbers, there is reduction in the temporal growth rates compared to the unladen flow. Chan et al.\cite{chan_2008} studied the stability of particle-laden round jet using a $tan{h}$ profile for the axial velocity with the shear layer to jet diameter as a fluid parameter. Particle mass loading parameter $Z$ (which is the product of the volume fraction and density ratio) and Stokes number were treated as particle parameters. They concluded that for small values of $Z$, first helical mode ($m=1$) is more unstable than the axi-symmetric mode ($m=0$), while for $Z>0.1$, $m=0$, is more unstable than $m=1$.

The study of particle-laden swirling jet in combustors has received a lot of interest in the recent years. Liu et al. \cite{Liu_etal_2021}, \cite{Liu_etal_2021_irregular} and Apte et al.\cite{sv_apte_2003} performed simulations of particle- laden, swirling flow in a coaxial jet combustor. In their simulations, the primary jet is air laden with spherical, glass particles, while a swirling stream of air flows through the annulus. They used an incompressible, spatially filtered Navier Stokes equation solver based on unstructured grids to compute the turbulent gas-phase. Particles are treated as point sources of momentum and are two-way coupled via the Navier Stokes equations. 
Particles are tracked in a Lagrangian framework. Their simulations showed that heavier particles with larger inertia rapidly penetrate the central re-circulation bubble while smaller particles respond quickly to the changes in local velocities (accelerations/decelerations) of the gas phase. Four-way coupled simulations of Liu et al.\cite{liufour_2023} and Liu et al.\cite{liuhydrodynamic_2023} with  methane as the central jet (primary jet) and air as the coaxial jet (secondary swirling jet) found that having particles around the central methane jet is beneficial for the stabilization of the flame.

\begin{figure}[!ht]
	\centering
	\includegraphics[width=0.4\textwidth]{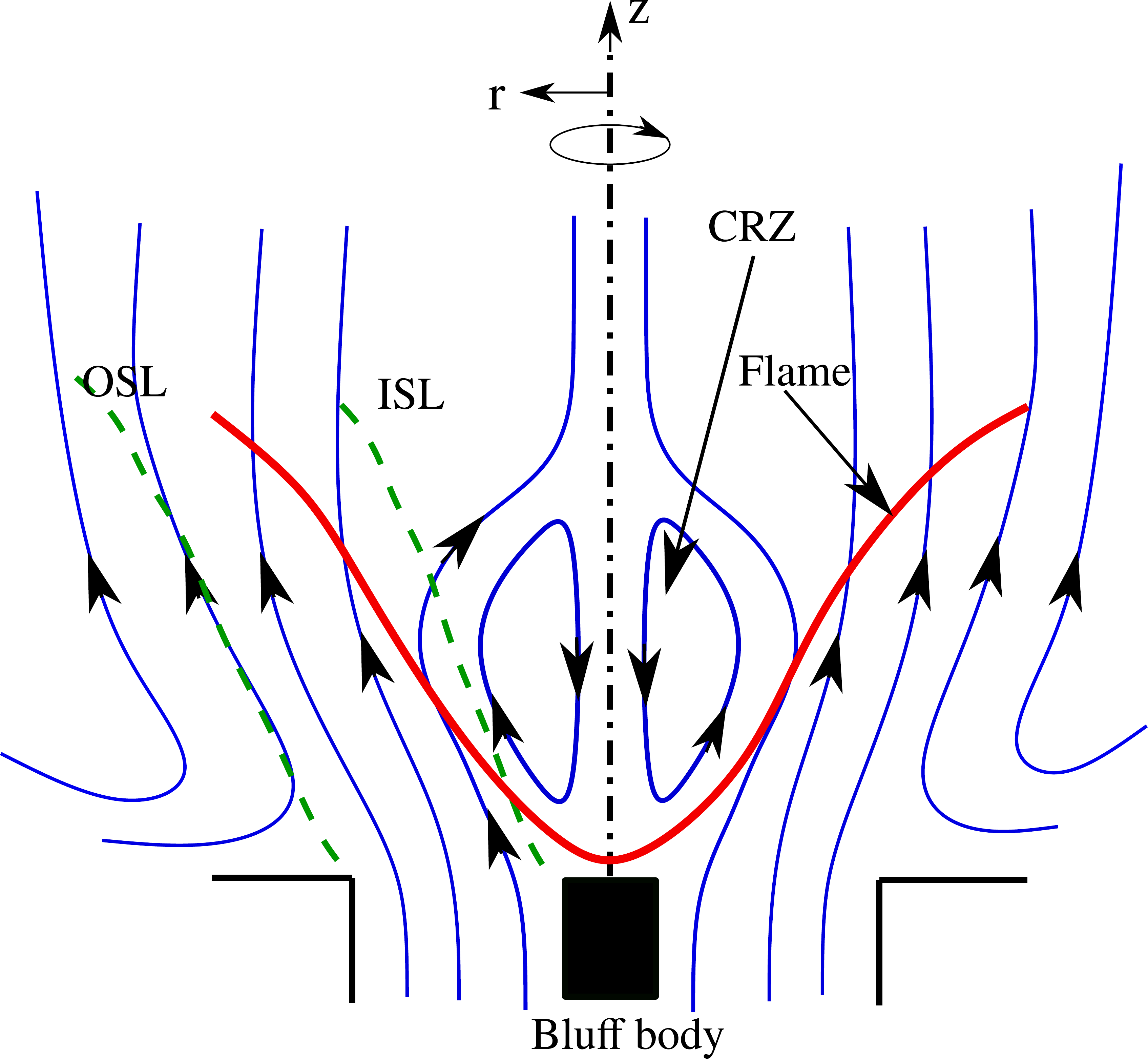}
	\caption{Schematic of the flow field in a typical annular swirl-stabilized combustor used in gas turbine engines. For sufficiently large ratios of axial fluxes of tangential momentum to axial momentum, vortex breakdown occurs resulting in the formation of a central re-circulation zone (CRZ). The shear layer between the annular jet and the ambient fluid is the outer shear layer (OSL), the shear layer between the annular jet and the CRZ is the inner shear layer (ISL). Figure adapted from Manoharan\cite{kiran_thesis_2019}.} 
	\label{fig:combustor_flowfield}  
\end{figure}

In the present work, we are interested in a different configuration of swirling jet in an annular swirl jet combustor typically used in gas turbine engines for aircraft propulsion. We will briefly look at the flow field in the combustor. Figure(\ref{fig:combustor_flowfield}) shows a schematic of the annular swirl jet combustor. Swirling jets are characterized by a non-dimensional parameter called a local swirl number which is defined as the ratio of the maximum azimuthal to maximum axial flow velocity ($S$) (Manoharan et al.\cite{kiran_2015}),  
\begin{equation}
S = \frac{U_{\theta,max}}{U_{z,max}}
\label{local_swirl_number_def}
\end{equation}                                                                                                                                                                                                 
%
	For sufficiently large swirl numbers, vortex breakdown occurs and a central re-circulation zone (CRZ) is formed. The shear layer between the annular jet and the surrounding fluid is termed the outer shear layer (OSL). The shear layer between the CRZ and the annular jet is termed the inner shear layer (ISL). Variation of azimuthal velocity as a function of radial coordinates establishes an azimuthal shear layer (ASL). The different shear layers mentioned above become unstable, thus resulting in an unsteady coherent flow structure called the Precessing Vortex Core (PVC). The PVC oscillates at a certain frequency and have been studied by Syred \cite{syred_2006}, Lieuwen \cite{lieuwen_2012} and  Bellows et al.\cite{Bellows_2007}.  Experimental studies on swirling jets by Liang and Maxworthy \cite{liangexperimental_2005}, Billant et al.\cite{billantexperimental_1998}, Leibovich \cite{leibovichstructure_1978}, Escuider \cite{escudiervortex_1988} and Gallaire et al.\cite{gallaire_ruith_2006} reveal that the breakdown of the vortex is strongly dependent on the swirl number. Previously several theoretical and experimental studies have described multiple instability mechanisms occurring in purely swirling flows (radial and axial velocities are zero but azimuthal velocity is non zero) and swirling jets (Gallaire and Chomaz \cite{gallaireinstability_2003}, Juniper\cite{juniperabsolute_2012} and Carnevale et al.\cite{carnevaleemergence_1994}). 
In the study of stability of swirling flows, Heaton \cite{Heaton_2007}, Le Diz{\`e}s et al.\cite{Fabre} considered Batchelor vortex (Batchelor\cite{batchelor1964axial}) (also referred to as \textit{q} vortex) as the baseflow velocity profile. They found a new set of unstable modes strongly driven by centrifugal instability, called centre modes, whose eigenmode shapes peaked in the vicinity of the vortex core. Heaton\cite{Heaton_2007} performed a high $Re$ asymptotic analysis and found that centre modes  become unstable if the quantity $H_{0}$,
\begin{equation}
H_{0}=2k\overline{W}_{r=0}\left(2k\left(\overline{W}\right)_{r=0}-m\left(\frac{d^{2}U_{z}}{dr^{2}}\right)_{r=0}\right) < 0.
\label{Heaton_criteria}
\end{equation}
where $\overline{W}_{r=0}=\left(\frac{dU_{\theta}}{dr}\right)_{0}$ and $\left(-\frac{d^{2}U_{z}}{dr^{2}}\right)_{r=0}$ are the baseflow axial vorticity and the derivative of the azimuthal vorticity at the jet centreline respectively. $k$ and $m$ are the axial and azimuthal wavenumbers respectively.  From Eq.(\ref{Heaton_criteria}) profiles with zero axial vorticity at the center of the vortex result in $H_{0}=0$ for which the centre modes remain stable. Also note that from Eq.(\ref{Heaton_criteria}) for axi-symmetric mode $m=0$, $H_{0}=4k^{2}\overline{W}_{\theta}^{2} > 0$. Therefore, only co-rotating disturbances exhibit unstable centre modes. Manoharan et al.\cite{kiran_2015} considered profiles with $H_{0}>0$, where the centre modes are stable, found that for uniform as well as variable density swirling jets, the only unstable modes are the shear layer modes. In a later work, Manoharan \cite{kiran_thesis_2019} reported that for this profile, the shear layer modes consisted of two types of modes called sinuous and varicose modes, where the eigenmode shapes peaked in the shear layers and not in the vortex core. Using Batchelor vortex as a baseflow profile, Le Diz{\`e}s and Fabre \cite{Fabre_TCFD} reported the presence of ring modes which are unstable and their eigenmode shapes largely confined to the shear layers. The Batchelor vortex also consists of continuous spectra (algebraic modes) as reported by Mao and Sherwin \cite{mao2012transient}. They performed non-modal analysis of the Batchelor vortex and showed that the continuous spectra (CS) showed a large transient growth indicative of a high degree of non-orthogonality of the continuous spectra modes. Optimal perturbation studies revealed the emergence of streamwise and azimuthal velocity streaks and the contamination of the vortex core. 
In the present work, we use the base flow profile closely following Oberleithner et al.\cite{oberleithner_2011} which was later modified by Manoharan \cite{kiran_thesis_2019}. The base flow profile of Oberleithner et al.\cite{oberleithner_2011} is an experimental fit to the time-averaged profiles obtained experimentally for an annular swirling jet configuration shown in figure(\ref{fig:combustor_flowfield}). The various shear layer thickness parameters (viz OSL, ISL, ASL) and the non-dimensional parameters namely, swirl number and backflow parameter(strength of the reverse flow in the CRZ) can be conveniently varied to simulate different operating conditions of the combustor. Manoharan et al.\cite{kiran_2015}, Manoharan \cite{kiran_thesis_2019} used this profile to understand the various instability mechanisms that contribute to the dynamics of the flow.  
We mentioned that the shear layers and the re-circulation zones are formed post-vortex breakdown which occurs at a certain critical swirl number. However, typical swirl numbers in gas turbine combustors at operating conditions are much higher than the critical swirl numbers at which vortex breakdown occurs. According to Manoharan et al.\cite{kiran_etal_2020} the unsteady PVC oscillation observed is due to instability associated with the post-vortex breakdown flow.  Temporal analysis by Manoharan \cite{kiran_thesis_2019} revealed the presence of three unstable modes namely, centre modes, sinuous and varicose modes. The hydrodynamic effect of the presence of the flame was modeled as a density gradient. Manoharan \cite{kiran_thesis_2019} for $m=1$, showed an increase in peak temporal growth rate with the addition of density gradient compared with the constant density case and attributed this to the fact that, the fluctuating baroclinic torque is in-phase with the source term contribution from rearrangement, stretching and tilting of base flow vorticity resulting in an increase in the net fluctuating vorticity generation rate. Their spatio-temporal calculations revealed that the self-sustained PVC oscillations observed in constant density swirling jets is the manifestation of $m = 1$ helical mode becoming globally unstable due to the presence of large regions of absolute instability in the flow field. 
\newline
In the present work, we consider the linear stability problem of the particle-laden swirling jet in the dilute suspension limit, in an annular combustor using the modified baseflow by Manoharan \cite{kiran_thesis_2019} which is based on the work of Oberleithner et al.\cite{oberleithner_2011}. However, we do not account for the density gradient caused by the presence of the flame. Here we consider the scenario of constant density of the continuous phase (gas phase). In the context of annular swirl-stabilized combustor, this would  correspond to the density ratio of burnt to unburnt gas equal to unity ($\gamma=1$) in Manoharan et al.\cite{kiran_2015} and Manoharan  \cite{kiran_thesis_2019}.  Recently Sharif et al.\cite{sharif2022} performed experiments on swirling dense sand-water jets in stagnant water and observed a drop in momentum flux along the recirculation zone due to the formation of the precessing vortex core further downstream. However, in the context of an annular combustor, the study of the effect of particles from a locally parallel linear stability perspective has not been reported in the literature. We are interested in the effect of particles on the centre modes. Although recent stability calculations of the unladen swirling jet profile by Manoharan \cite{kiran_thesis_2019} has revealed that the first helical mode ($m=1$) is responsible for the centre mode instability, linear stability calculations of the particle laden swirling jet for the above combustor configuration has not been reported in the literature.
 
 We study the effect of the baseflow particle concentration with uniform as well as non-uniform profiles. Comparisons from the unladen case of Manoharan \cite{kiran_thesis_2019} are made and the effect on the temporal growth rates at different Stokes number (ratio of particle relaxation time to the convective time scale of the flow), local swirl number and backflow parameter (non-dimensional reverse flow velocity magnitude at the jet centerline) are studied. Useful insights regarding the stability of swirling jet from the present study can be gained which aid in understanding the role of particles in stabilizing the vortex core.

This paper is divided into the following sections.  Governing equations are introduced along with the non-dimensionalization of flow variables, linearized equations, normal mode form and boundary conditions are discussed in section (\ref{sec:formulation}). Linearized vorticity transport equations are discussed in (\ref{sec:Linearised vorticity dynamics}). Base flow details and validation cases are presented in section (\ref{sec:validation and base state}). The results are discussed in section (\ref{sec: results and discussion}) and summarized in section (\ref{sec:conclusion}).

\section{Governing equations}
\label{sec:formulation} 
We use volume averaged equations to convert the Lagrangian description for particles into an Eulerian formulation resulting in a volume fraction field by associating an averaging volume whose dimension ($l_{avg}$) is much greater than the inter-particle distance but far smaller compared to the length scale ($L_{c}$) over which macroscopic changes occur (the characteristic length scale of the flow). The size of the particle is far smaller compared to the inter-particle distance implying dilute suspension (see figure 1 in Warrier et al.\cite{warrier2022linear}). The governing equations are integrated over the averaging volume. The equations in polar cylindrical coordinates used here are derived following the procedure of volume average described by Chu and Prosperetti\cite{CHU_2016} and Warrier et al.\cite{warrier2022linear}   
Mass conservation equation for the $j^{th}$ phase is given by,
\begin{equation}
\frac{\partial\alpha^{j}}{\partial t}+\nabla\cdot\alpha^{j}\left\langle \mathbf{u}^{j}\right\rangle^{(j)} =0.
\label{chap:5:eq:averaged_mass_conservation}
\end{equation}

Mass conservation for the fluid phase ($j=1$, $\alpha^{(1)}= 1-\alpha, \langle\mathbf{u}^{\left(1\right)}\rangle^{\left(1\right)}=\mathbf{u}$) is given by,
\begin{equation}
\frac{\partial\left(1-\alpha\right)}{\partial t}+\nabla\cdot\left(\left(1-\alpha\right)\mathbf{u}\right)=0.
\label{Eq:fluid_mass_continuity}
\end{equation}

Mass conservation for the particle phase ($j=2$, $\alpha^{(2)}= \alpha, \langle\mathbf{u}^{\left(2\right)}\rangle^{\left(2\right)}=\mathbf{u}_{p}$) is given by,
\begin{equation}
\frac{\partial\alpha}{\partial t}+\nabla\cdot\left(\alpha\mathbf{u}_{p}\right)=0.
\label{Eq:particle_mass_continuity}
\end{equation}

The non dimensional volume averaged momentum equation for the continuous phase is given by, 

\begin{equation}
\frac{\partial}{\partial t}\left(\left(1-\alpha\right)\mathbf{u}\right)+\nabla\cdot\left(\left(1-\alpha\right)\mathbf{u}\mathbf{u}\right)=-\left(1-\alpha\right)\nabla p + \left(1-\alpha\right)\mu\nabla\cdot\left[\nabla\overline{\mathbf{u}}+\left(\nabla\overline{\mathbf{u}}\right)^{T}\right]-\frac{\gamma}{St}\alpha\left(\mathbf{u}-\mathbf{u}_{p}\right).
\label{chap:5:eq:averaged_fluid_momentum}
\end{equation}
where $\mathbf{u}$, $\mathbf{u}_{p}$ are the volume averaged gas and particle velocities respectively. $\gamma$ is the particle to gas density ratio, $\alpha$ is the particle volume fraction, $\overline{\mathbf{u}}=(1-\alpha)\mathbf{u} + \alpha\mathbf{u}_{p}$, $St$ is the Stokes number defined as the ratio of particle relaxation time to the flow convective time, $\mu$ is the fluid viscosity. The particle momentum equation is given by,
\begin{equation}
\frac{\partial}{\partial t}\left(\alpha\mathbf{u}_{p}\right)+\nabla\cdot\left(\alpha\mathbf{u}_{p}\mathbf{u}_{p}\right)=\frac{1}{St}\alpha\left(\mathbf{u}-\mathbf{u}_{p}\right).
\label{chap:5:eq:averaged_particle_momentum}
\end{equation}
We take special care for the pressure gradient term in the volume averaged momentum equation in the case of swirling jets. From \cite{CHU_2016}, the pressure gradient term for the $j^{th}$ phase is given by,
\begin{equation}
	\alpha^{j}\nabla\overline{p}=\alpha^{j}\nabla\left((1-\alpha)\left\langle p^{1}\right\rangle + \alpha\left\langle p^{2}\right\rangle\right).
	\label{eq:meanpressure}
\end{equation} 
where $\left\langle p^{1}\right\rangle$ and $\left\langle p^{2}\right\rangle$ are the volume averaged fluid and particle pressure respectively. Recent investigations of swirling gas-particle jet in a coaxial jet combustor (Liu et al.\cite{Liu_etal_2021} and Liu et al.\cite{Liu_etal_2021_irregular}) show that particle pressure is modified in dense (concentrated) particle suspensions where particle-particle collisions lead to energy exchange between the gas and particles (Gidaspow\cite{Gidaspow} and Koch\cite{Koch}) which modify the particle pressure. These effects are ignored in the present work as there are no particle-particle collisions in the dilute suspension regime. Considering particles to be millimeter sized, the volume concentration assuming the number density to be $O(10^{5}/m^{3})$ (sufficient number for averaging) turns out to be,

	\begin{equation}
	\alpha=\frac{NV_{i}}{V}= Nd_{p}^{3}/L^{3}= 10^{5}(10^{-3})^{3}\sim 10^{-4}.
	\end{equation} 
	where $n=N/L^{3}$ is the number density of particles and $d_{p}$ is the diameter of each particle. Therefore we limit the volume concentration considered in this work to $10^{-4}$ i.e, one part in ten thousand to ensure dilute suspension. Ignoring particle-particle collisions that modify the particle pressure, we have $\left\langle p^{1}\right\rangle=\left\langle p^{2}\right\rangle=p$.
 Eq.(\ref{eq:meanpressure}) now becomes,
\begin{equation}
\alpha^{j}\nabla\overline{p}=\alpha^{j}\nabla p.
\end{equation} 
which is what appears in Eq.(\ref{chap:5:eq:averaged_fluid_momentum}). Also note that the pressure gradient term does not appear in the particle phase equations in the non dimensional form Eq.(\ref{chap:5:eq:averaged_particle_momentum}), owing to large particle to gas density ratios as discussed in our previous work on particle laden planar jet (Warrier et al.\cite{warrier2022linear}).
 The radial component of the velocity is $u_{r}$, azimuthal component is $u_{\theta}$ and the axial component is $u_{z}$. The volume averaged equations for the fluid and dispersed phases are given below. The non-dimensional radial component equation for the fluid phase is given by,


The non dimensional radial component equation for the fluid phase is given by,
\begin{equation*}
\left(1-\alpha\right)\frac{\partial u_{r}}{\partial t}+\left(1-\alpha\right)u_{r}\frac{\partial u_{r}}{\partial r}+\left(1-\alpha\right)u_{z}\frac{\partial u_{r}}{\partial z}+\frac{\left(1-\alpha\right)}{r}u_{\theta}\frac{\partial u_{r}}{\partial\theta} -\frac{\left(1-\alpha\right)}{r}u_{\theta}^{2}=-\left(1-\alpha\right)\frac{\partial p}{\partial r}+
\end{equation*}

\begin{equation}
=\frac{\left(1-\alpha\right)}{Re}\left\{ \nabla^{2}\left(\left(1-\alpha\right)u_{r}\right)-\frac{\left(1-\alpha\right)u_{r}}{r^{2}}-\frac{2}{r^{2}}\frac{\partial}{\partial\theta}\left(\left(1-\alpha\right)u_{\theta}\right)\right\} -\frac{\alpha\gamma}{St}\left(u_{r}-u_{pr}\right).
\label{chap:5:eq:non_linear_radial_momentum}
\end{equation}

The non dimensional azimuthal component equation for the fluid phase is given by,
\begin{equation*}
	\left(1-\alpha\right)\frac{\partial u_{\theta}}{\partial t}+\left(1-\alpha\right)u_{r}\frac{\partial u_{\theta}}{\partial r}+\frac{\left(1-\alpha\right)}{r}u_{\theta}\frac{\partial u_{\theta}}{\partial\theta}+\left(1-\alpha\right)u_{z}\frac{\partial u_{\theta}}{\partial z}+\frac{\left(1-\alpha\right)}{r}u_{r}u_{\theta}=\frac{-\left(1-\alpha\right)}{r}\frac{\partial p}{\partial\theta}
\end{equation*}

\begin{equation}
=\frac{-\left(1-\alpha\right)}{r}\frac{\partial p}{\partial\theta}+\frac{\left(1-\alpha\right)}{Re}\left\{ \nabla^{2}\left(\left(1-\alpha\right)u_{\theta}\right)-\frac{\left(1-\alpha\right)u_{\theta}}{r^{2}}+\frac{2}{r^{2}}\frac{\partial}{\partial\theta}\left(\left(1-\alpha\right)u_{r}\right)\right\} -\frac{\alpha\gamma}{St}\left(u_{\theta}-u_{p\theta}\right).
\label{chap:5:eq:non_linear_azimuthal_momentum}
\end{equation}

The non dimensional axial component equation for the fluid phase is given by,
\begin{equation}
\left(1-\alpha\right)\frac{\partial u_{z}}{\partial t}+\left(1-\alpha\right)u_{r}\frac{\partial u_{z}}{\partial r}+\frac{1}{r}\left(1-\alpha\right)u_{\theta}\frac{\partial u_{z}}{\partial\theta}+\left(1-\alpha\right)u_{z}\frac{\partial u_{z}}{\partial z}=-\left(1-\alpha\right)\frac{\partial p}{\partial z}+\frac{\left(1-\alpha\right)}{Re}\left(\nabla^{2}\left(\left(1-\alpha\right)u_{z}\right)\right)-\frac{\alpha\gamma}{St}\left(u_{z}-u_{pz}\right).
\label{chap:5:eq:non_linear_axial_momentum}
\end{equation}
 where $\nabla^{2}=\left(\frac{\partial^{2}}{\partial r^{2}}+\frac{1}{r}\frac{\partial}{\partial r}+\frac{1}{r^{2}}\frac{\partial^{2}}{\partial\theta^{2}}+\frac{\partial^{2}}{\partial z^{2}}\right)$. 

\subsection*{Particle phase equations}
The non dimensional momentum equation for the particle phase is given by,
\begin{equation}
\frac{\partial}{\partial t}\left(\alpha\mathbf{u}_{p}\right)+\nabla\cdot\left(\alpha\mathbf{u}_{p}\mathbf{u}_{p}\right)=\frac{1}{St}\alpha\left(\mathbf{u}-\mathbf{u}_{p}\right).
\label{chap:5:eq:particle_momentum}
\end{equation}

The non dimensional radial momentum equation for particle phase is given by,
\begin{equation}
\frac{\partial u_{pr}}{\partial t}+u_{pr}\frac{\partial u_{pr}}{\partial r}+\frac{u_{p\theta}}{r}\frac{\partial u_{pr}}{\partial\theta}+u_{pz}\frac{\partial u_{pr}}{\partial z}-\frac{u_{p\theta}^{2}}{r}=\frac{1}{St}\left(u_{r}-u_{pr}\right).
\label{chap:5:eq:particle_radial_momentum}
\end{equation}
The non dimensional azimuthal momentum equation for particle phase is given by,
\begin{equation}
\begin{aligned}
    \frac{\partial u_{p\theta}}{\partial t}+u_{pr}\frac{\partial u_{p\theta}}{\partial r}+u_{pz}\frac{\partial u_{p\theta}}{\partial z}+\frac{u_{p\theta}}{r}\frac{\partial u_{p\theta}}{\partial\theta}+\frac{u_{pr}u_{p\theta}}{r}=\frac{1}{St}\left(u_{\theta}-u_{p\theta}\right).
\end{aligned}
\label{chap:5:eq:particle_azimuthal_momentum}
\end{equation}
The non dimensional axial momentum equation for particle phase is given by,
\begin{equation}
\frac{\partial u_{pz}}{\partial t}+u_{pr}\frac{\partial u_{pz}}{\partial r}+\frac{u_{p\theta}}{r}\frac{\partial u_{pz}}{\partial\theta}+u_{pz}\frac{\partial u_{pz}}{\partial z}=\frac{1}{St}\left(u_{z}-u_{pz}\right).
\label{chap:5:eq:axial_radial_momentum}
\end{equation}
In order to perform local analysis, we linearize the governing equation (volume averaged Navier Stokes equation) around a base state that is assumed to be steady and locally parallel. We consider the  base state to be locally parallel swirling jet. Following the assumption of Saffman\cite{saffman_1962}, the mean particle velocity is assumed to be equal to mean fluid velocity.  We write the total flow as a perturbation series with the leading order term as the base state and perturbations of $O(\epsilon)$,
\begin{equation}
\mathbf{q}\left(r,\theta,z,t\right) =\mathbf{Q}(r) + \epsilon \mathbf{q}^{'}\left(r,\theta,z,t\right) + O(\epsilon)^2.
\label{total flow}
\end{equation} 
where the base flow quantity $\mathbf{Q}$ is given by, 
\begin{equation}
\mathbf{Q}(r) =  \left[U_{r},U_{\theta},U_{z},P,U_{pr},U_{\theta},U_{pz},\Lambda\right]^{T}.
\label{chap:5:lsa}
\end{equation}
where $U_{pr}$, $U_{p\theta}$ and  $U_{pz}$ are the particle radial, azimuthal and axial base state velocities. $\Lambda$ is the baseflow concentration field. For a locally parallel, zero inertia baseflow we have, 
\begin{equation}
	\begin{aligned}
	U_{r} = U_{pr} = 0,\\
	U_{\theta}= U_{p\theta}= U_{\theta}(r),\\
	U_{z}=U_{pz}=U_{z}(r), \\
	\Lambda = \Lambda(r).
	\end{aligned}
     \label{chap:5:eq:jet_baseflow}
\end{equation}
 Base state pressure is a function of the radial coordinate and is given by $P(r)=\int\frac{U_{\theta}^{2}}{r}dr$.
Eq.(\ref{total flow}) is substituted in the mass conservation equation Eq.(\ref{chap:5:eq:averaged_mass_conservation}) with $j=1,2$ for the fluid and particle phase respectively, the volume averaged momentum equations Eqs.(\ref{chap:5:eq:non_linear_radial_momentum} - \ref{chap:5:eq:axial_radial_momentum}). The linearized equations are obtained by considering the $O(\epsilon)$ terms. The disturbance quantities (primed quantities) are further assumed to have a travelling wave form solution given by,  
\begin{equation}
\mathbf{q}^{'}\left(r,\theta,z,t\right)= \tilde{\mathbf{q}}\left(r\right)e^{i\left(kz+m\theta-\omega t\right)}
\label{chap:5:normal mode form equations}
\end{equation}
 $\tilde{\mathbf{q}}\left(r\right)=\left[\tilde{u}_{r}\left(r\right),\tilde{u}_{\theta}\left(r\right),\tilde{u}_{z}\left(r\right),\tilde{p}\left(r\right),\tilde{u}_{pr}\left(r\right),\tilde{u}_{p\theta}\left(r\right),\tilde{u}_{pz}\left(r\right),\tilde{\alpha}\left(r\right)\right]^{T}$ is the complex eigenvector,  $\omega$ is the complex eigenvalue. The wavenumbers $k$ and $m$ are real. The radial direction is in-homogeneous while the axial and azimuthal directions are homogeneous and periodic respectively. In temporal analysis, disturbance is localised in space and grows exponentially in time. The wavenumbers in the axial direction $k$ and the azimuthal direction $m$ are real while temporal frequency $\omega$ is complex. Eq.(\ref{chap:5:normal mode form equations}) is substituted in the linearized volume averaged equations  to obtain the normal mode equations. The normal mode form of the equations in the matrix form is given in Appendix \ref{appendix_a}.


At infinity we impose the far field boundary condition. As $r\rightarrow\infty$, all perturbations vanish,
\begin{equation}
\begin{Bmatrix}\tilde{u}_{r} & \tilde{u}_{\theta} & \tilde{u}_{z} & \tilde{p} & \tilde{u}_{pr} & \tilde{u}_{p\theta} & \tilde{u}_{pz} & \tilde{\alpha}\end{Bmatrix}^{T}=0.
\end{equation}
%
The centreline condition for single phase jet has been derived previously by  Batchelor and Gill \cite{batchelor_gill_1962}, later by Khorrami et al.\cite{khorrami_1989}. Here we derive the centreline condition for the particle disturbance quantities. The underlying idea is the same as that for the single phase equations, that at the centreline, the volume averaged particle velocity and the volume fraction has to have a vanishing dependence on the azimuthal direction for which,
\begin{equation}
\begin{array}{cccc}
\underset{r\rightarrow0}{lim}\frac{\partial\boldsymbol{u}^{'}}{\partial\theta}=0, & \underset{r\rightarrow0}{lim}\frac{\partial\boldsymbol{u_{p}}^{'}}{\partial\theta}=0, & \underset{r\rightarrow0}{lim}\frac{\partial p^{'}}{\partial\theta}=0, & \underset{r\rightarrow0}{lim}\frac{\partial\alpha^{'}}{\partial\theta}=0\end{array}.
\label{eq:centreline_constraints}
\end{equation}
Fluid perturbation quantities have to obey the following conditions at the centreline,
\begin{equation}
\underset{r\rightarrow0}{lim}\frac{\partial\boldsymbol{u}^{'}}{\partial\theta}=\left(\frac{\partial u_{r}^{'}}{\partial\theta}-u_{\theta}^{'}\right)\boldsymbol{e_{r}}+\left(\frac{\partial u_{\theta}^{'}}{\partial\theta}+u_{r}^{'}\right)\boldsymbol{e_{\theta}}+\left(\frac{\partial u_{z}^{'}}{\partial\theta}\right)\boldsymbol{e_{z}}=0. 
\label{r0_perturb_fluid_cond}
\end{equation}
Substituting the normal mode form for the perturbation in Eq.(\ref{r0_perturb_fluid_cond}), we get,
\begin{equation}
\label{centreline_fluid}
\begin{aligned}
im\tilde{u}_{r}-\tilde{u}_{\theta}=0, \\
im\tilde{u}_{\theta}+\tilde{u}_{r}=0 ,\\
im\tilde{u}_{z}=0,  \\
\underset{r\rightarrow0}{lim}\frac{\partial p^{'}}{\partial\theta}=im\tilde{p}=0. \\   
\end{aligned}
\end{equation}
while the constraints on the particle perturbation quantities at the centreline (from Eq.(\ref{eq:centreline_constraints}))are the following,
\begin{equation}
    \left(\frac{\partial u_{pr}^{'}}{\partial\theta}-u_{p\theta}^{'}\right)\boldsymbol{e_{r}}+\left(\frac{\partial u_{p\theta}^{'}}{\partial\theta}+u_{pr}^{'}\right)\boldsymbol{e_{\theta}}+\left(\frac{\partial u_{pz}^{'}}{\partial\theta}\right)\boldsymbol{e_{z}}=0.
\label{r0_perturb_particle_cond}
\end{equation}
Substituting the normal mode form for the perturbations in Eq.(\ref{r0_perturb_particle_cond}), we get,
\begin{equation}
\label{centreline_particle}
\begin{aligned}
im\tilde{u}_{pr}-\tilde{u}_{p\theta}=0, \\
im\tilde{u}_{p\theta}+\tilde{u}_{pr}=0, \\
im\tilde{u}_{pz}=0,  \\
\underset{r\rightarrow0}{lim}\frac{\partial\alpha^{'}}{\partial\theta}=im\tilde{\alpha}=0. \\   
\end{aligned}
\end{equation}
It is clear from Eq.(\ref{centreline_fluid}) and Eq.(\ref{centreline_particle}) that the centreline conditions depends upon the azimuthal wavenumber of the perturbation.
\subsubsection{Axi-symmetric perturbations, $m=0$}
The centreline condition for the axi-symmetric perturbation ($m=0$) for fluid and particle field is obtained
from Eq.(\ref{centreline_fluid}) and Eq.(\ref{centreline_particle}) respectively as,
\begin{equation}
\label{m0_bc}
\begin{aligned}
\tilde{u}_{r}=0, \tilde{u}_{\theta}=0, \\
\tilde{u}_{z} , \tilde{p} \mbox{ must be finite} \\
\tilde{u}_{pr}=0,  \tilde{u}_{p\theta}=0, \\
\tilde{u}_{pz} , \tilde{\alpha} \mbox{ must be finite} \\
\end{aligned}
\end{equation}
\subsubsection{First helical mode, $m=\pm1$} 
From Eq.(\ref{centreline_fluid}) and Eq.(\ref{centreline_particle}), we derive the conditions for the first helical disturbance. Here $m=1$ denotes helical disturbance in clockwise direction, while $m=-1$ denotes denotes disturbances in the anticlockwise sense with respect to the direction of swirl in the baseflow.
\begin{equation}
\begin{aligned}
\tilde{u}_{r} \pm i\tilde{u}_{\theta}=0, \\
\tilde{u}_{r} \pm i\tilde{u}_{\theta}=0, \\
\tilde{u}_{z}=0, \\
\tilde{p}=0. \\
\end{aligned}
\label{m1_r0_condition_on_fluid}
\end{equation}
similarly for the perturbed particle fields,
\begin{equation}
\begin{aligned}
\tilde{u}_{pr} \pm i\tilde{u}_{p\theta}=0, \\
\tilde{u}_{pr} \pm i\tilde{u}_{p\theta}=0, \\
\tilde{u}_{pz}=0, \\
\tilde{\alpha}=0. \\
\end{aligned}
\label{m1_r0_condition_on_particles}
\end{equation}
Now for $\left|m\right|=1$, you have two linearly dependent conditions for the fluid and particle fields. The other condition is obtained by evaluating the fluid and particle continuity equation as $r\rightarrow0$,
\begin{equation}
\left(2\left(1-\Lambda\right)\frac{d\tilde{u}_{r}}{dr}\right)_{r=0}-2\left(\frac{d\Lambda}{dr}\tilde{u}_{r}\right)_{r=0} +\left(im\left(1-\Lambda\right)\frac{d\tilde{u}_{\theta}}{dr}\right)_{r=0} -\left(im\tilde{u}_{\theta}\frac{d\Lambda}{dr}\right)_{r=0}=0.
\label{m1_r0_VPL_fluid_continuity}
\end{equation}
Using continuity equation for the particle phase,
\begin{equation}
\left(\frac{d\Lambda}{dr}\tilde{u}_{pr}\right)_{r=0}+2\left(\Lambda\frac{d\tilde{u}_{pr}}{dr}\right)_{r=0}+\left(\tilde{u}_{pr}\frac{d\Lambda}{dr}\right)_{r=0}+  \left(im\Lambda\frac{d\tilde{u}_{p\theta}}{dr}\right)_{r=0} +\left(im\tilde{u}_{p\theta}\frac{d\Lambda}{dr}\right)_{r=0}=0.
\label{m1_r0_VPL_particle_continuity}
\end{equation}
For uniform particle loading (particle concentration, $\Lambda$ is a constant), the jet centreline condition for $m=\pm1$ becomes,
\begin{equation}
2\left(\frac{d\tilde{u}_{r}}{dr}\right)_{r=0}+im\left(\frac{d\tilde{u}_{\theta}}{dr}\right)_{r=0}=0.
\label{m1_r0_CPL_fluid_continuity}
\end{equation}
\begin{equation}
2\left(\frac{d\tilde{u}_{pr}}{dr}\right)_{r=0}+im\left(\frac{d\tilde{u}_{p\theta}}{dr}\right)_{r=0}=0.
\label{m1_r0_CPL_particle_continuity}
\end{equation}
Eq.(\ref{m1_r0_CPL_fluid_continuity}) is the same as the centreline condition for single phase flow. The centreline condition for the particle velocity field is given by Eq.(\ref{m1_r0_CPL_particle_continuity}) which is similar to  fluid's except that the fluid velocity components are replaced by particle velocity components. Notice that for non-uniform particle concentration profiles, Eq.(\ref{m1_r0_VPL_fluid_continuity},\ref{m1_r0_VPL_particle_continuity}) tells us that the centreline conditions depend on the baseflow concentration profile and its derivative. For Gaussian concentration profile, Eq.(\ref{m1_r0_VPL_fluid_continuity},\ref{m1_r0_VPL_particle_continuity}) reduce to Eq.(\ref{m1_r0_CPL_fluid_continuity}, \ref{m1_r0_CPL_particle_continuity}).
\subsubsection{Higher helical disturbances $|m|>1$}
For higher helical disturbances with $|m|>1$ (clockwise and anticlockwise sense of rotation), we have from Eq.(\ref{centreline_fluid}) and Eq.(\ref{centreline_particle}),
\begin{equation}
\label{mg1_bc_fluid}
\begin{aligned}
\tilde{u}_{r}  = \tilde{u}_{\theta} = 0, \\
\tilde{u}_{z}  = \tilde{p} = 0,  \\
\tilde{u}_{pr} = \tilde{u}_{p\theta} = 0, \\
\tilde{u}_{pz} = \tilde{\alpha} = 0. \\
\end{aligned}
\end{equation}
The system of linearized Navier--Stokes equation in the normal form posed as an eigenvalue problem can be written in the matrix form as
\begin{equation}
	A\tilde{\mathbf{q}}=\omega B \tilde{\mathbf{q}}.
	\label{normal_mode_swirling_jet}
\end{equation}

where $A$ and $B$ are operators which is a function of the base state and the wave numbers $k$ and $m$. The temporal eigenvalue is given by $\omega$. The dispersion relation is numerically solved by using a pseudospectral collocation technique employing Chebyshev polynomials (see Boyd \cite{Boyd_2013} for details). The eigenspectrum for a given wavenumber ($k$) is computed using the $QZ$ algorithm (Saad\cite{saad_2011}) by using MATLAB's \textit{eig} function. To track an unstable eigenmode for a range of wavenumbers, we use the eigen value obtained by the $QZ$ algorithm for a given wavenumber $k$ as a guess eigen value and compute eigen value at wavenumber $k+\delta k$ (For all calculations in the present work, $\delta k=0.005$ is found to be sufficient) using Arnoldi iteration. (using the \textit{eigs} function in MATLAB). Arnoldi iteration computes a small number of eigen values only around the guess value making it computationally less expensive (Saad\cite{saad_2011}).
Similar procedure is employed by Warrier et al.\cite{warrier2022linear} for the tracking of unstable modes in particle-laden planar jets.
For particle-laden round jet, the domain is defined from $r=0$ to $r\rightarrow\infty$. We use the mapping function used to transform from $[0, \infty]$ to $\eta\in[-1, 1]$ given by Khorrami et al.\cite{khorrami_1989},

\begin{equation}
r=f\left(\eta\right)=a\left(\frac{1+\eta}{b-\eta}\right).
\end{equation}
where $a$ and $b$ are related as,
\begin{equation}
b=1+\frac{2a}{r_{max}}.
\label{khorami_b_parameter}
\end{equation}
The parameter $a$ concentrates most of the points between $r=0$ and $r=a$. We take $a=3$ as suggested by Khorrami et al. \cite{khorrami_1989}. The derivatives in the transformed physical domain are related to the computational domain as,
\begin{equation}
	\begin{aligned}
	\frac{d}{dr}=\frac{1}{f'}\frac{d}{d\eta}, \\
	\frac{d^{2}}{dr^{2}}=\frac{1}{f'^{2}}\frac{d^{2}}{d\eta^{2}}-\left(\frac{f''}{f'^{3}}\right)\frac{d}{d\eta}.
	\end{aligned}
\end{equation}
where $\frac{d}{d\eta}$ and $\frac{d^{2}}{d\eta^{2}}$ are the Chebyshev differentiation matrices.
\section{Linearized vorticity dynamics}
\label{sec:Linearised vorticity dynamics}
The effect of particle addition on the linearized vorticity budget is studied by deriving the linearized vorticity transport equations in cylindrical coordinates. The terms contributing to the net generation rate of vorticity fluctuations are identified and evaluated for the base flow variation used in the present study and their corresponding hydrodynamic eigenmodes. 
Manoharan et al.\cite{kiran_2015} and later Manoharan\cite{kiran_thesis_2019} used vorticity transport equations for the particle free swirling jet and identified source terms. We follow the same procedure to derive the linearized vorticity budget equations for particle-laden swirling jets for uniform particle loading. The linearized vorticity budget equations are obtained by taking curl of the linearized volume averaged equations.
Radial component of the vorticity budget equation for the fluid phase is given by,
\begin{equation}
\underset{\frac{D\Omega_{r}^{'}}{Dt}}{\underbrace{\frac{\partial\Omega_{r}^{'}}{\partial t}+\frac{U_{\theta}}{r}\frac{\partial\Omega_{r}^{'}}{\partial\theta}+U_{z}\frac{\partial\Omega_{r}^{'}}{\partial z}}}=\underset{\circled{1}}{\underbrace{-\frac{im\tilde{u}_{r}}{r}\frac{dU_{z}}{dr}}}+\underset{\circled{2}}{\underbrace{\left(\frac{dU_{\theta}}{dr}+\frac{U_{\theta}}{r}\right)ik\tilde{u}_{r}}}+\underset{\circled{3}}{\underbrace{\frac{1}{Re}\left(\frac{1}{r}\frac{\partial\mathcal{F}_{z}^{'}}{\partial\theta}-\frac{\partial\mbox{\ensuremath{\mathcal{F}_{\theta}^{'}}}}{\partial z}\right)}}-\underset{\circled{4}}{\underbrace{\frac{\gamma\Lambda}{St}\left(\Omega_{r}^{'}-\Omega_{pr}^{'}\right)}}.
\label{eq:radial_vorticity_budget}
\end{equation}
Azimuthal component of the vorticity budget equation for the fluid phase is given by,

\[
\underset{\frac{D\Omega_{\theta}^{'}}{Dt}}{\underbrace{\frac{\partial\Omega_{\theta}^{'}}{\partial t}+\frac{U_{\theta}}{r}\frac{\partial\Omega_{\theta}^{'}}{\partial\theta}+U_{z}\frac{\partial\Omega_{\theta}^{'}}{\partial z}}}=\underset{\circled{5}}{\underbrace{\left(\frac{d^{2}U_{z}}{dr^{2}}+\frac{dU_{z}}{dr}\frac{d}{dr}\right)\tilde{u}_{r}}}+\underset{\circled{6}}{\underbrace{ik\tilde{u}_{z}\frac{dU_{z}}{dr}}}+\underset{\circled{7}}{\underbrace{\left(\frac{dU_{\theta}}{dr}-\frac{U_{\theta}}{r}\right)\frac{im\tilde{u}_{z}}{r}}}+\underset{\circled{8}}{\underbrace{\frac{2ikU_{\theta}}{r}\tilde{u}_{\theta}}}
\]

\begin{equation}
+\underset{\circled{9}}{\underbrace{\frac{1}{Re}\left(\frac{\partial\mathcal{F}_{r}^{'}}{\partial z}-\frac{\partial\mathcal{F}_{z}^{'}}{\partial r}\right)}}
-\underset{\circled{10}}{\underbrace{\frac{\gamma\Lambda}{St}\left(\Omega_{\theta}^{'}-\Omega_{p\theta}^{'}\right)}}+\underset{\circled{11}}{\underbrace{\left(u_{z}^{'}-u_{pz}^{'}\right)\frac{\gamma}{St}\frac{d\Lambda}{dr}}}.
\label{eq:azimthal_vorticity_budget}
\end{equation}

Axial component of the vorticity budget equation for the fluid phase is given by,

\[
\underset{\frac{D\Omega_{z}^{'}}{Dt}}{\underbrace{\frac{\partial\Omega_{z}^{'}}{\partial t}+\frac{U_{\theta}}{r}\frac{\partial\Omega_{z}^{'}}{\partial\theta}+U_{z}\frac{\partial\Omega_{z}^{'}}{\partial z}}} =\underset{\circled{12}}{\underbrace{\left(\frac{d^{2}U_{\theta}}{dr^{2}}+\frac{1}{r}\frac{dU_{\theta}}{dr}-\frac{U_{\theta}}{r^{2}}\right)\tilde{u}_{r}}}+\underset{\circled{13}}{\underbrace{ik\left(\frac{dU_{\theta}}{dr}+\frac{U_{\theta}}{r}\right)\tilde{u}_{z}}}-\underset{\circled{14}}{\underbrace{ik\frac{dU_{z}}{dr}\tilde{u}_{\theta}}}
\]
\begin{equation}
+\underset{\circled{15}}{\underbrace{\frac{1}{Re}\left(\frac{1}{r}\left(\frac{\partial}{\partial r}\left(r\mathcal{F}_{\theta}^{'}\right)-\frac{\partial\mbox{\ensuremath{\mathcal{F}_{r}^{'}}}}{\partial\theta}\right)\right)}}-\underset{\circled{16}}{\underbrace{\frac{\gamma\Lambda}{St}\left(\Omega_{z}^{'}-\Omega_{pz}^{'}\right)}}-\underset{\circled{17}}{\underbrace{\left(u_{\theta}^{'}-u_{p\theta}^{'}\right)\frac{\gamma}{St}\left(\frac{d\Lambda}{dr}\right)}}.
\label{eq:axial_vorticity_budget}
\end{equation}

The subscripts $p$ indicate the vorticity components for the particle phase. The Eqs.(\ref{eq:radial_vorticity_budget},\ref{eq:azimthal_vorticity_budget},\ref{eq:axial_vorticity_budget}) represent the component-wise rate of change of perturbation vorticity following the fluid element which has contributions from tilting and stretching of the vortex lines, viscous dissipation of the vorticity and the additional terms due addition of particles and gradients in the baseflow particle concentration. The terms $\circled{4}$, $\circled{10}$, $\circled{11}$, $\circled{16}$ and $\circled{17}$ represent the curl of the two-way coupling term. To see this consider the curl of the two-way coupling term in the vector form, 
\begin{equation}
\nabla \times \frac{\gamma\Lambda}{St}\left(\mathbf{u'} - \mathbf{u_{p}'}\right) = \frac{\gamma\Lambda}{St}\left(\mathbf{\Omega'} - \mathbf{\Omega_{p}'}\right) + \frac{\gamma}{St}\left(\nabla \Lambda \times \left(\mathbf{u'} - \mathbf{u_{p}'}\right)\right).
\label{curl_two_way_coupled_term}
\end{equation}
The first term in the R.H.S of Eq.(\ref{curl_two_way_coupled_term}) given by $\left(\frac{\gamma\Lambda}{St}\left(\mathbf{\Omega'} - \mathbf{\Omega_{p}'}\right)\right)$ is the difference in the fluid and particle perturbation vorticities, given by the  terms $\circled{4}$, $\circled{10}$ and $\circled{16}$ in the vorticity budget equation. The second term in Eq.(\ref{curl_two_way_coupled_term}) given by $\left(\frac{\gamma}{St}\left(\nabla \Lambda \times \left(\mathbf{u'} - \mathbf{u_{p}'}\right)\right)\right)$ is the term that is non-zero when the baseflow particle concentration gradient is misaligned with the perturbation fluid - particle relative velocity vector (slip velocity).
These terms are given by  $\circled{11}$ and $\circled{17}$ in the vorticity budget equations.
This term is identically zero in the case of uniform particle concentration or when the baseflow particle concentration is such that the baseflow concentration gradient is aligned with the slip velocity vector.  The terms $\circled{3}$, $\circled{9}$ and $\circled{15}$ represent the viscous dissipation of vorticity. 
The other source terms in the vorticity budget equation represents contributions to the net generation rate of vorticity fluctuations from rearrangement of base flow vorticity by velocity disturbances, unsteady vortex stretching due to base flow velocity gradients. In typical gas turbine combustors, jet Reynolds number is of the order of few thousands. In the present work we consider $Re=10000$ and therefore neglect the viscous dissipation of vorticity while computing the various budget terms, however viscous terms in the linearized equations are taken into account while solving the eigen value problem.

\section{Validation and base state}
\label{sec:validation and base state}
For validation purpose, we reduce the formulation to the equations to the single phase flow (unladen flow) by switching off the terms corresponding to the particle phase and comparing it against the eigen spectrum of round jet by Gallaire and Chomaz \cite{gallaire_chomaz_2003}. Both eigenspectra show a good match. For the case of swirling jet in an annular combustor, we compare against the particle-free eigenspectra of Manoharan\cite{kiran_thesis_2019}. See Appendix \ref{appendix_a} for the matrix form of the equations.
\subsection{Validation with unladen non swirling round jet: Gallaire and Chomaz profile} 
We consider the case of swirling jet proposed by Gallaire and Chomaz \cite{gallaire_chomaz_2003}. The velocity field is given by the expressions,
\begin{equation}
\begin{aligned}
U_{z}(r) = a + \frac{1}{1+(e^{(r^{2}log 2)}-1)^{N}}, \\
U_{r}(r) = 0, \\
U_{\theta}(r) =0.
\end{aligned}
\label{gallaire_chomaz_swirl_equation}
\end{equation}
where the non-dimensional shear layer thickness parameter $N=3$. The advection parameter $a$ is the free-stream axial velocity measured from the axis.

\begin{figure}[!ht]
	\centering
	\includegraphics[width=0.45\textwidth]{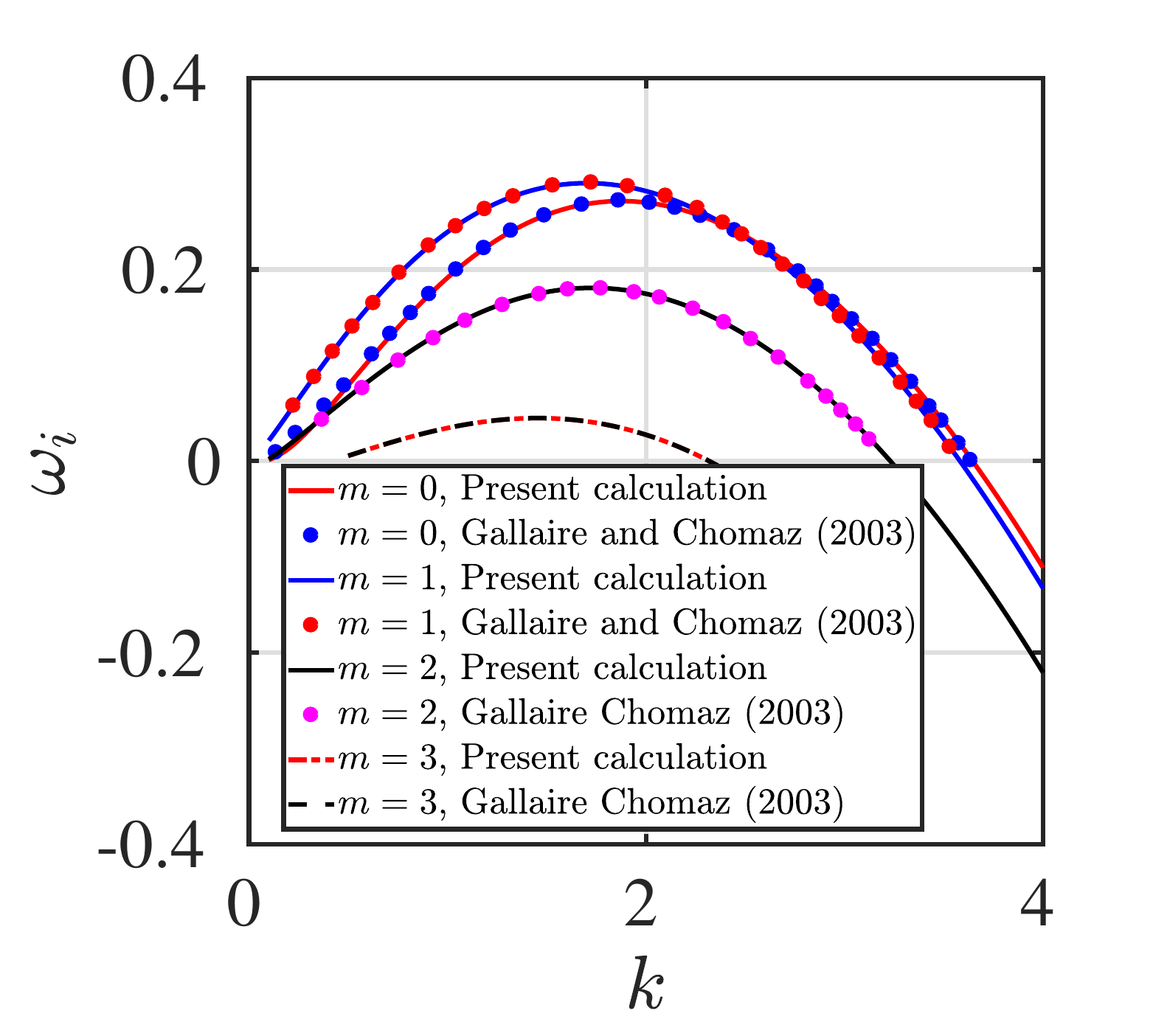}
	\caption{Shows the agreement of the temporal growth rates of the unladen non swirling jet with that of Gallaire and Chomaz \cite{gallaire_chomaz_2003}. Particle terms are switched off and only the fluid terms are retained and validated. }
	\label{unladen_non_swirling_gallaire_chomaz}  
\end{figure}

Figure \ref{unladen_non_swirling_gallaire_chomaz} shows the dispersion relation of the unladen non swirling round jet obtained by switching off the particle terms (i.e, putting base state and perturbation particle concentration and particle velocities to zero). We see a good agreement with the growth rates given by Gallaire Chomaz \cite{gallaire_chomaz_2003} for $N=3$, $r_{v}=0.9$, $a=0$, $q=0$, $\alpha=4$  when the system of equations are reduced to single phase flow. Since the base state has no swirl, positive and negative helical modes ($\pm m$) are the same. The first helical mode $m=1$ has the highest growth rate compared to axi-symmetric and higher azimuthal modes. 
In the present validation case with the Gallaire and Chomaz profile \cite{gallaire_chomaz_2003}, the non dimensional shear layer thickness (non-dimensionalised by the jet core radius) is $0.13$ for $N=3$ (Eq.(\ref{gallaire_chomaz_swirl_equation})) for which $m=1$ mode is most unstable. 

\subsection{Base state: Locally parallel swirling jet profile in an annular combustor (Manoharan \cite{kiran_thesis_2019}). 
\label{sec:Baseflow_validation}}

\begin{figure*}[!ht]
    \includegraphics[width=0.4\textwidth]{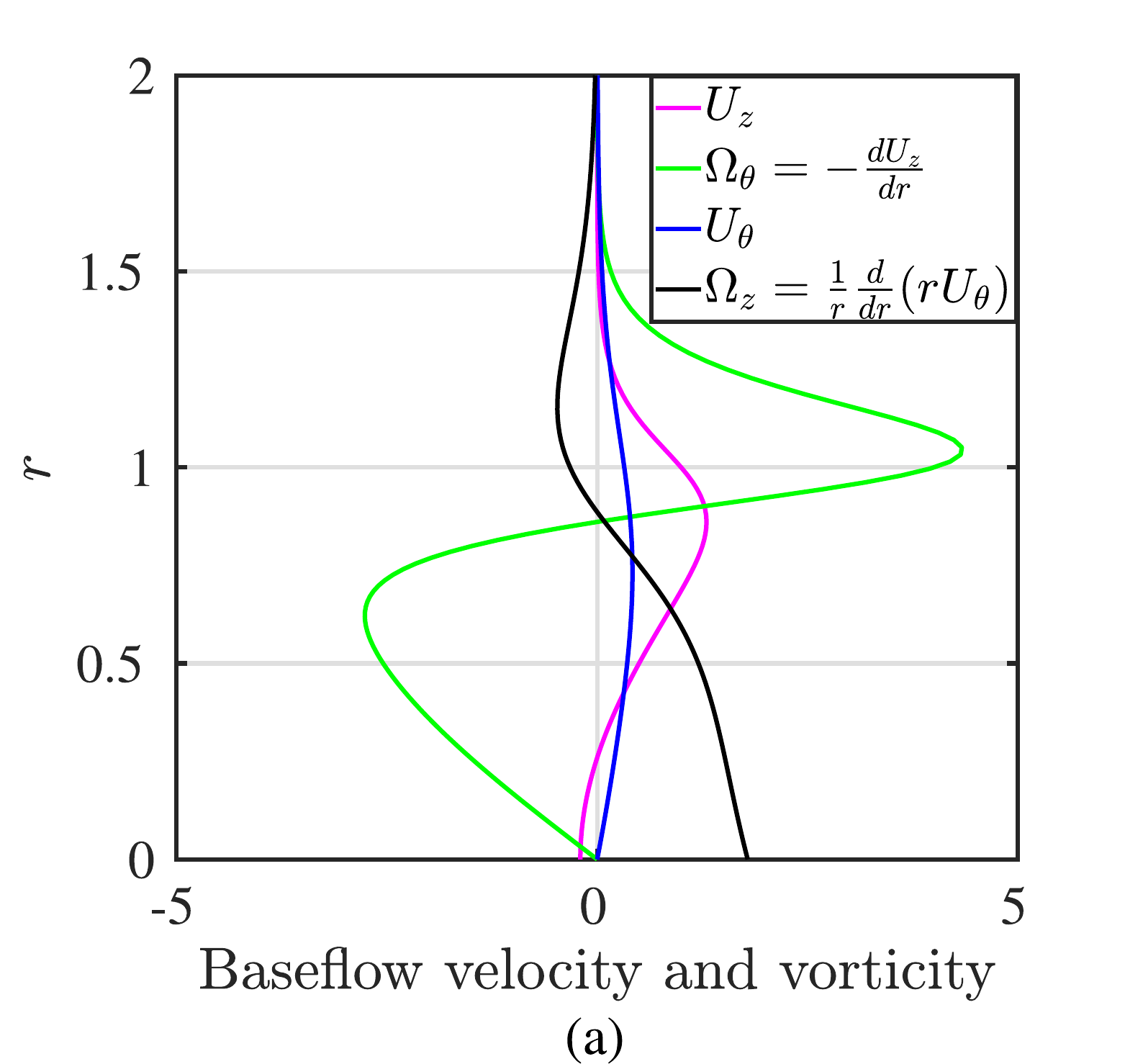}
	\includegraphics[width=0.4\textwidth]{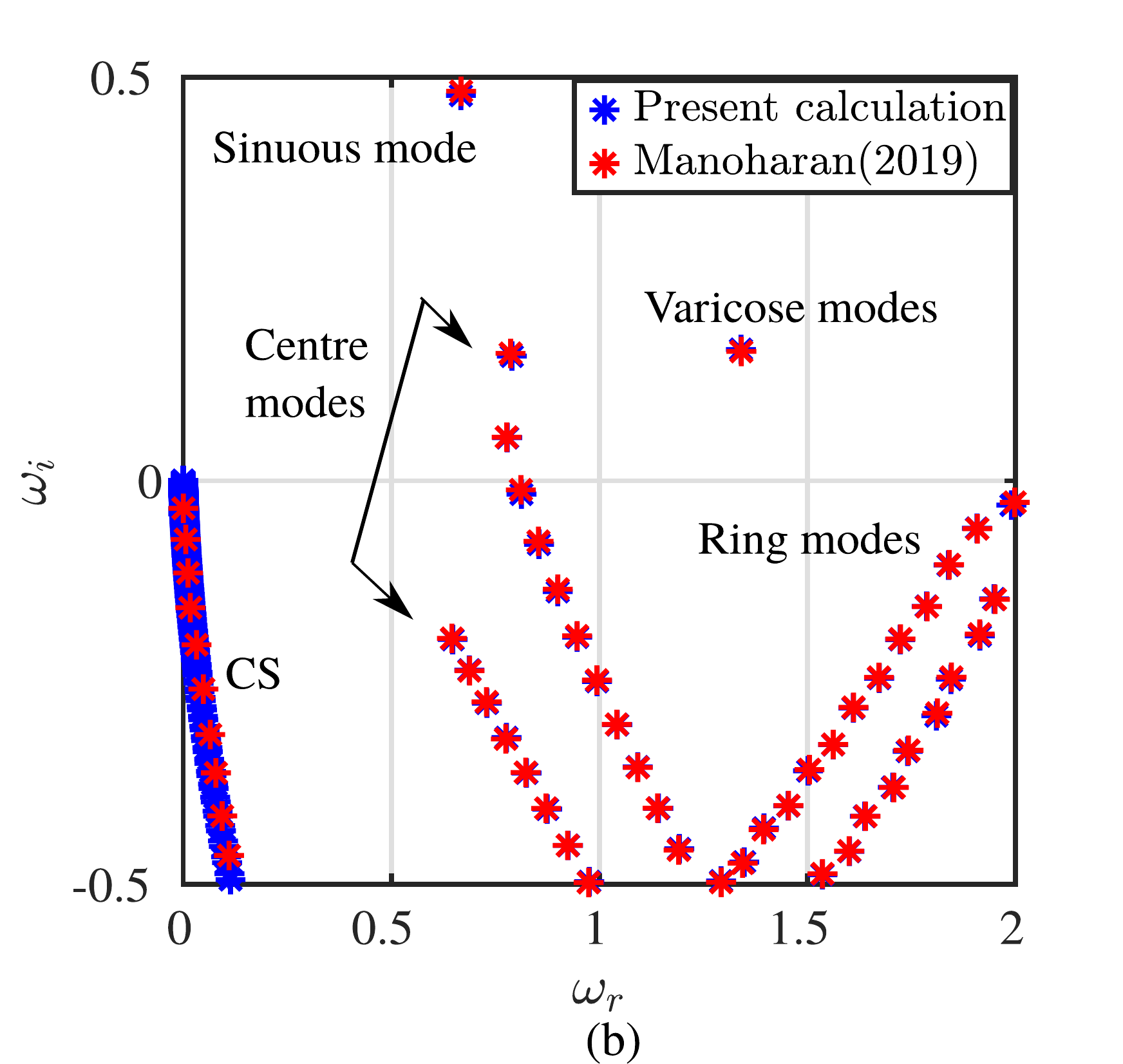} 
	\caption{(a) shows the base flow axial, azimuthal velocity and the vorticity components. $U_{z}$ is the axial velocity and $\Omega_{z}$ is the axial vorticity component which is non zero at the centreline while $\Omega_{\theta}$ is the azimuthal vorticity component. (b) Eigen spectra of the unladen swirling jet showing various unstable modes  at  $Re=10000$. CS stands for continuous spectrum modes. For the profile shown, local swirl number $S=0.5$, backflow parameter $\beta=0.2$, the shear layer thickness and fitting parameters are tabulated in table \ref{table:oberliethner_details}.}
	\label{fig:unladen_oberliethner_basestate}  
\end{figure*}

The baseflow used to study the stability of particle-laden swirling jet is that of a locally parallel velocity field set up in an annular swirl jet combustor which is schematically shown in figure(\ref{fig:combustor_flowfield}). The local profile has been used previously by Oberleithner et al.\cite{oberleithner_2011}, Manoharan et al.\cite{kiran_2015}, Manoharan\cite{kiran_thesis_2019}.  As mentioned in the introduction section of the paper, at sufficiently large swirl numbers, bubble type vortex breakdown occurs (Billant et al.\cite{billantexperimental_1998}) and this creates a re-circulation zone in the vortex core. This also results in the formation of multiple shear layers (as opposed to having a single shear layer in other swirling flow base states, given for instance by Gallaire and Chomaz\cite{gallaire_chomaz_2003}, Gallaire et al.\cite{gallaire_ruith_2006},\cite{gallaireinstability_2003}). The multiple shear layers formed are the inner shear layer formed between the annular jet and CRZ, outer shear layer formed between the annular jet and the ambient flow, an azimuthal shear layer due to the gradients in the azimuthal velocity profile. The shear layer parameters used in the present work is that considered by Manoharan \cite{kiran_thesis_2019} shown in table (\ref{table:oberliethner_details}).

 \begin{equation}
  \begin{aligned}
    U_{r} = 0, \\
    U_{\theta} = 4SF_{3}\left(r\right)\left(1-F_{4}\left(r\right)\right), \\
   	U_{z} = 4BF_{1}\left(r\right)\left(1-BF_{2}\left(r\right)\right), \\  	
  \end{aligned}
  \label{eq:oberliethner_base}
 \end{equation}
where, 
\begin{equation}
	F_{j}\left(r\right)=\frac{1}{\left[1 + \left(e^{r^{2}b_{j}} - 1\right)^{N_{j}}\right]},
\end{equation} 
where $j=1, 2, 3, 4$.  The parameter $B$ is given by the expression,
\begin{equation}
	B = \left[1 + \left(1+\beta\right)^{1/2}\right], 
\end{equation}
The parameter $\beta$ is the non dimensional reverse flow axial velocity magnitude at the jet centreline, $\beta=-U_{z}\left(r=0\right)$. The $N_{j}$'s are the shear layer thickness parameters and the $b_{j}$'s are the experimental fitting parameters. The radial velocity is zero since the profiles are locally parallel. Figure(\ref{fig:unladen_oberliethner_basestate}) shows the azimuthal and axial velocity and vorticity field as a function of the radial distance from the jet centreline.

\begin{table}[h]
\centering
	\caption{Shows the values of axial and azimuthal velocity shear layer thickness parameters along with their corresponding fitting parameters used in the present study(identical to \cite{kiran_thesis_2019}).}
 \label{oberliethner_details}
	\begin{tabular}{@{}lll@{}}
		\toprule
		Thickness parameter & Symbol & value \\
		\midrule
\hline
Outer axial shear layer & $N_{1}$ & $4$\\
\hline
Inner axial shear layer & $N_{2}$ & $1$\\
\hline
Outer azimuthal shear layer & $N_{3}$ & $2$\\
\hline
Inner azimuthal shear layer  & $N_{4}$ & $0.5$\\
\hline
Outer axial shear layer fitting parameter  & $b_{1}$ & $0.693$\\
\hline
Inner axial shear layer fitting parameter  & $b_{2}$ & $0.693$\\
\hline
Outer azimuthal shear layer fitting parameter  & $b_{3}$ & $0.69$\\
\hline
Outer azimuthal shear layer fitting parameter  & $b_{4}$ & $0.2$\\
\hline
		\bottomrule
	\end{tabular}
\label{table:oberliethner_details}
\end{table}

\begin{table}[h]
		\caption{Shows the grid convergence for the most unstable mode (centre mode, sinuous mode and varicose mode) for the unladen and particle laden case at $St=1$, $k=1.2$ for $N=300,400,500$ }
	\label{table:grid_convergence}
	\begin{tabular}{@{}lllll@{}}
		\toprule
	 Swirling jet & $N$ & centre mode $(\omega)$ & sinuous mode $(\omega)$& varicose mode $(\omega)$ \\
\hline
Unladen swirling jet & $300$ & $0.789653401 + 0.1545032069i$ &$0.66720880855+ 0.4776152284i$ &$1.3401030476+ 0.162735945i$\\
			 & $400$ & $0.789653401 + 0.1545032069i$ & $0.66720880855 + 0.4776152284i$ & $1.3401030476+0.162735945i$\\
			&$500$& $0.789653401+0.1545032069i$ & $0.66720880855+0.4776152284i$ & $1.3401030476+0.162735945i$  \\
			\hline
			Particle-laden   & $300$  &$0.79237+0.114521i$ & $0.6688122010+0.441368710774i$ &$1.3478+0.1238i$\\
			swirling jet & $400$ & $0.79237+0.114521i$ & $0.6688122010+0.441368710775i$ & $1.3475+0.1239i$ \\
			$St=1,\Lambda=10^{-4}$& $500$ & $0.79237+0.114521i$ & $0.6688122010+0.441368710777i$ & $1.3475+0.1239i$\\	
\hline
		\bottomrule
	\end{tabular}
\label{table:grid_convergence}
\end{table}

  Based on the large $Re$ asymptotics of Heaton \cite{Heaton_2007} and Le Diz{\`e}s and Fabre\cite{Fabre}, swirling jets have an additional set of modes termed as centre modes which become unstable  if the parameter $H_{0}$ in Eq.(\ref{Heaton_criteria}) is less than zero. Eq.(\ref{Heaton_criteria}) highlight  the importance of the baseflow azimuthal and axial vorticity, especially at the jet centreline. 
For a locally parallel swirling flow, the radial vorticity is zero, the azimuthal vorticity and the axial vorticity is given by,

\begin{equation}
	\begin{aligned}
	\Omega_{\theta}=-\frac{dU_{z}}{dr}, \\
	\Omega_{z}=\frac{1}{r}\frac{d}{dr}\left(rU_{\theta}\right).
	\end{aligned}
\label{eq:omega_z_baseflow}
\end{equation}

At the jet centreline, axial vorticity is given by $\Omega_{z}\left(0\right)=2\left(\frac{dU_{\theta}}{dr}\right)_{0}$ (by L'H$\hat{o}$pital rule on Eq.(\ref{eq:omega_z_baseflow})).
We can rewrite Heaton's Criteria Eq.(\ref{Heaton_criteria}) as,
\begin{equation}
H_{0}=\left(k\Omega_{z}(0)\right)^{2} + \left(km\Omega_{z}(0)\left(\frac{d\Omega_{\theta}}{dr}\right)_{0}\right) < 0. 
\label{Heaton_criteria_rearranged}
\end{equation}

It is clear from Eq.(\ref{Heaton_criteria_rearranged}) that the baseflow values at the jet centreline play a crucial role in the instability of the centre mode. Generic swirling jet profiles of Gallaire and Chomaz \cite{gallaire_chomaz_2003}, \cite{gallaireinstability_2003} have $\Omega_{z}(0)=0$ which makes $H_{0}=0$ from Eq.(\ref{Heaton_criteria_rearranged}) and do not exhibit centre modes. We note from Eq.(\ref{Heaton_criteria_rearranged}) that axi-symmetric modes and counter-rotating helical modes are stable. We therefore consider only co-rotating helical disturbances. For the baseflow considered in the present study $H_{0} <0$. The presence of a centre bluff body (used for stabilizing the flame) can significantly alter the value of the axial vorticity at the centreline. For the baseflow parameters originally used by Oberleithner \cite{oberleithner_2011}, it was found that the centre modes (driven by centrifugal instability) remain stable. Manoharan \cite{kiran_thesis_2019} modified the parameter values for which, by the use of Eq.(\ref{Heaton_criteria}), the centre modes are unstable (see table(\ref{table:oberliethner_details})). As swirl number increases, the axial vorticity at the centreline also increases and this is expected to play a vital role in the instability of the centre mode.
\subsection{Base state for particle-laden swirling jet}
 As for the base state for the particle phase, we assume that the base flow particle velocity is equal to the baseflow fluid velocity. We investigate both uniform and non uniform particle concentration profiles and limit the magnitude of the particle concentration to $10^{-4}$ and vary the Stokes number $St=10^{-5}$ to $1$. While the assumption of the particle baseflow velocity being equal to the baseflow fluid velocity is appropriate for low Stokes numbers, however for $St=1$, the particles may not faithfully follow the fluid streamline. In the context of planar mixing layers, for particles with $St=1$, Eaton and Fessler \cite{EATON_1994} showed that high density particles in the dilute limit exhibit preferential concentration where the particles migrate from the vortex core of the rolled up shear layer to a region of  high strain rate and low vorticity (i.e, the braid region). In the present work in order to study the linear stability characteristics of the above annular combustor configuration, we assume that for $St=1$, particle mean velocity is equal to the fluid mean velocity. This however needs to be compared with experiments and numerical simulations to asses the validity of the assumption and to obtain a more realistic fluid-particle baseflow.   

\section{Results and discussion}
\label{sec: results and discussion}

We now discuss the stability of a swirling jet in an annular swirl-stabilized combustor. Figure (\ref{fig:unladen_oberliethner_basestate}a,b) shows the locally parallel base-state profile and the temporal spectra of the unladen swirling jet obtained by switching-off the particle terms and the particle phase equations compared against the eigenspectra of Manoharan  \cite{kiran_thesis_2019}. The base flow parameters for this case are tabulated in table (\ref{table:oberliethner_details}). The jet Reynolds number is $10000$, which is typically seen in swirl combustors. 
We see that switching off the particle terms, the eigenspectra of the unladen jet is obtained which is in good agreement with  Manoharan \cite{kiran_thesis_2019}. 
	Table (\ref{table:grid_convergence}) shows the eigenvalue convergence for three different number of grid points ($N=300,400,500$). For the unladen laden jet, difference between the eigenvalues for $N=400,500$ is of the order of $10^{-10}$. For particle-laden jet, the difference between the eigenvalues for $N=400,500$ is of the order of $10^{-5}$ for the real part and $10^{-6}$ for the imaginary part of the eigenvalue of the centre mode. The eigenspectra ($\omega_{r} $ vs $\omega_{i}$) is calculated at wavenumber $k=1.2$, $St=1$ and $ \Lambda=10^{-4}$.
For $m=1$, with backflow parameter $\beta=0.2$ and local swirl number $S=0.5$, we see that the centre mode is unstable. Sinuous and varicose modes (shear layer modes) are unstable while the two sets of ring modes and the discrete equivalent of the continuous spectra are found to be stable at all wavenumbers. 
\subsection{Eigenspectra in the presence of particles}
\begin{figure*}[!ht]
	\centering
		\includegraphics[width=0.7\textwidth]{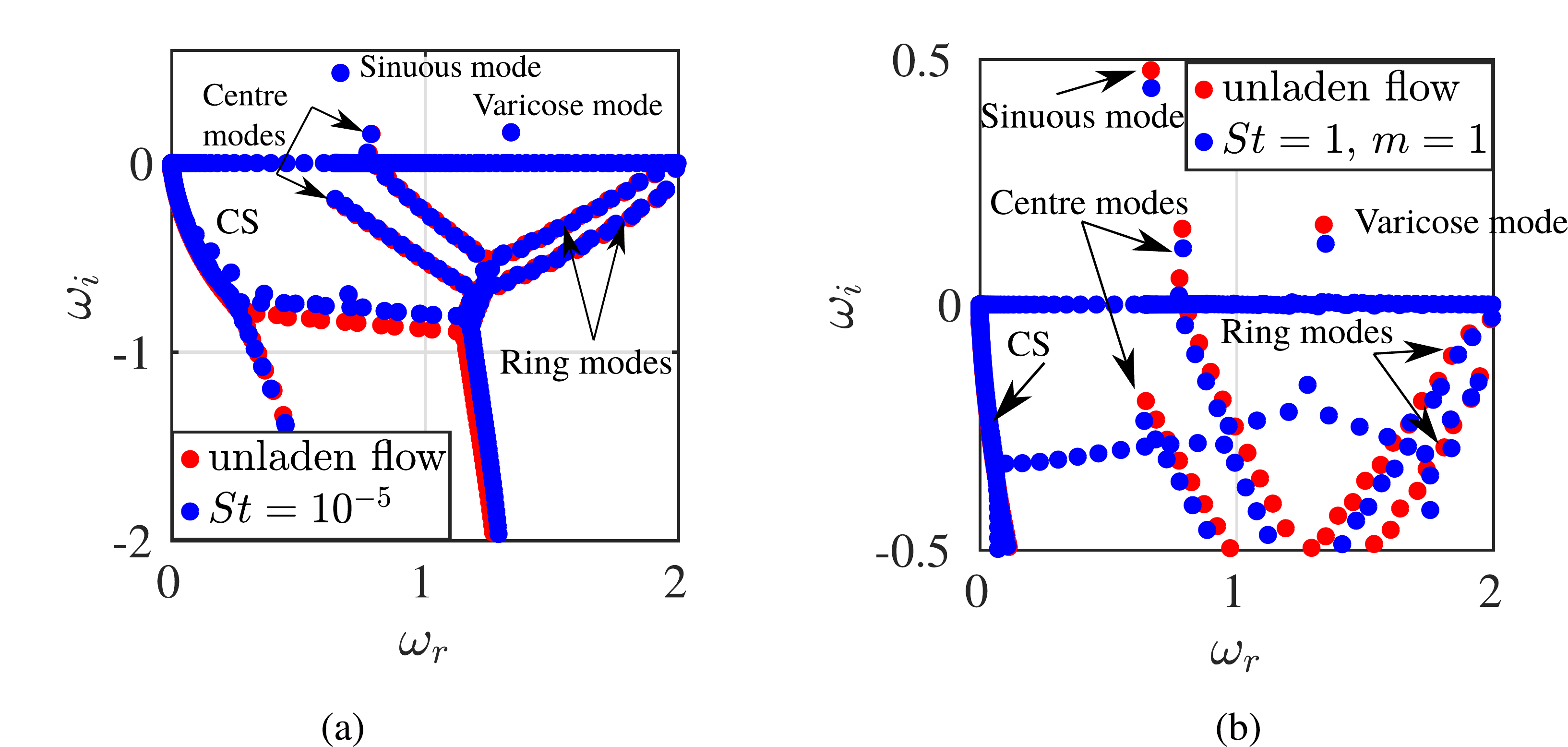}
	\caption{(a) Eigen spectra of particle-laden swirling jet at low Stokes number of $10^{-5}$ showing various unstable modes. (b) Eigen spectra at $St=1$  $Re=10000$, local swirl number $S=0.5$ and backflow parameter $\beta=0.2$ and $k=1.2$. CS stands for Continuous spectrum modes.}
	\label{low_stokes_oberleithner_spectra}  
\end{figure*}

\begin{figure}[!ht]
	\centering
		\includegraphics[width=0.7\textwidth]{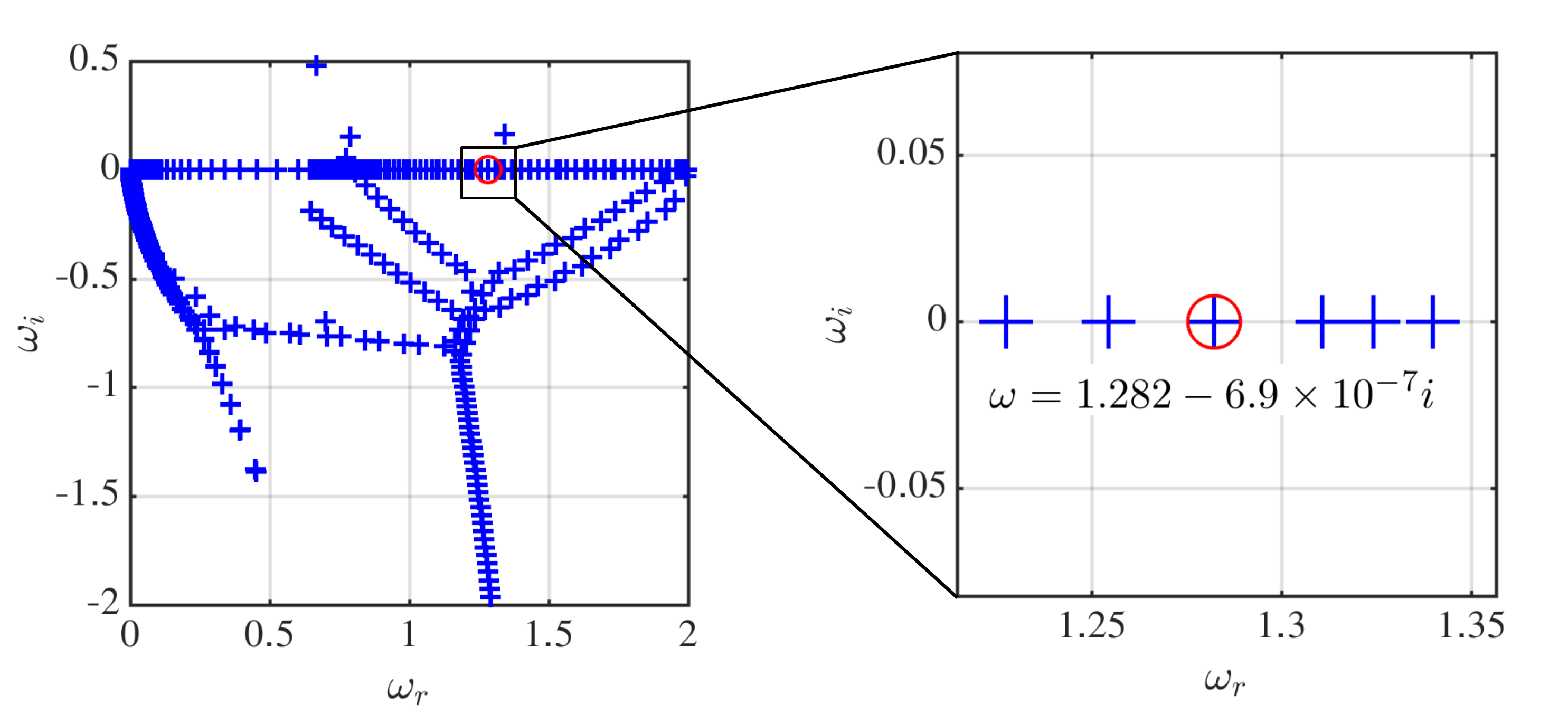}
	\caption{The eigenspectra of particle-laden swirling jet at $St=10^{-5},k=1.2,\Lambda=10^{-4}$. We consider the red "o" as one of the several modes that appear whose growth rate is small but negative. Such stable modes are also seen in the eigenspectra of particle-laden mixing layers, see figures (11b,d) by Narayanan and Lakehal\cite{Narayananetal2002}. }
	\label{fig:new_set_of_modes}  
\end{figure}

Figure (\ref{low_stokes_oberleithner_spectra}a) shows the eigenspectra of particle-laden jet at Stokes number of $St=10^{-5}$.  We note that the presence of particles (for the same set of baseflow parameters) brings in new set of modes which are absent in the unladen jet. These modes have very small growth rates as shown in figure(\ref{fig:new_set_of_modes}). The modes are decaying modes with small growth rates as shown in figure(\ref{fig:new_set_of_modes}). Such modes are also seen in the eigenspectra of particle-laden mixing layers reported by Narayanan and Lakehal\cite{Narayananetal2002}. For the unstable (growing) modes of particle-laden mixing layers and planar jet, it was observed that at low Stokes number, temporal growth rate of particle-laden flow is greater than that of the unladen case. This is attributed to the fact that at low Stokes number, the drag force due to particles is negligible when compared to the viscous dissipation term. Due to increase in mixture density, the effective kinematic viscosity decreases, increasing its dissipation. However, in the present case $Re=10000$, where most typical combustors operate, so that the difference in growth rates is negligible as seen in figure(\ref{low_stokes_oberleithner_spectra}a). As we increase Stokes number to unity, the effect of particle addition on the unstable mode centre and  shear layer modes (viz, sinuous and varicose modes) is decrease in the growth rate compared to that of the unladen swirling jet.
At $St=1$, for the same baseflow parameters and wavenumber ($m=1$, $k=1.2$), we see from figure (\ref{low_stokes_oberleithner_spectra}b) that, the discrete equivalent of the continuous spectra (CS) deviates from that of the unladen and the low Stokes number ($St=10^{-5}$) CS modes. These modes are also found in the case of unladen Batchelor vortex by Mao and Sherwin\cite{mao2012transient}. They found on performing non-modal analysis of the Batchelor vortex, that the continuous spectra showed a high degree of non orthogonality resulting in large transient growth. In the present baseflow of the swirling jet profile, we see that the continuous spectra eigenmodes remain stable for the unladen and particle-laden cases as well.  Further, a new branch of stable modes emerge from the CS which connects the stable part of the centre mode branch and all the way to the ring mode to the right of the spectra. The ring modes too seem to deviate from the usual two sets of stable modes seen in unladen spectra and the low Stokes number ($St=10^{-5}$) spectra. Both the CS and the ring modes remain stable for all Stokes numbers and wavenumbers. Among these unstable modes, we focus on the unstable centre modes below.

\begin{figure*}[!ht]
	\centering
\includegraphics[width=0.7\textwidth]{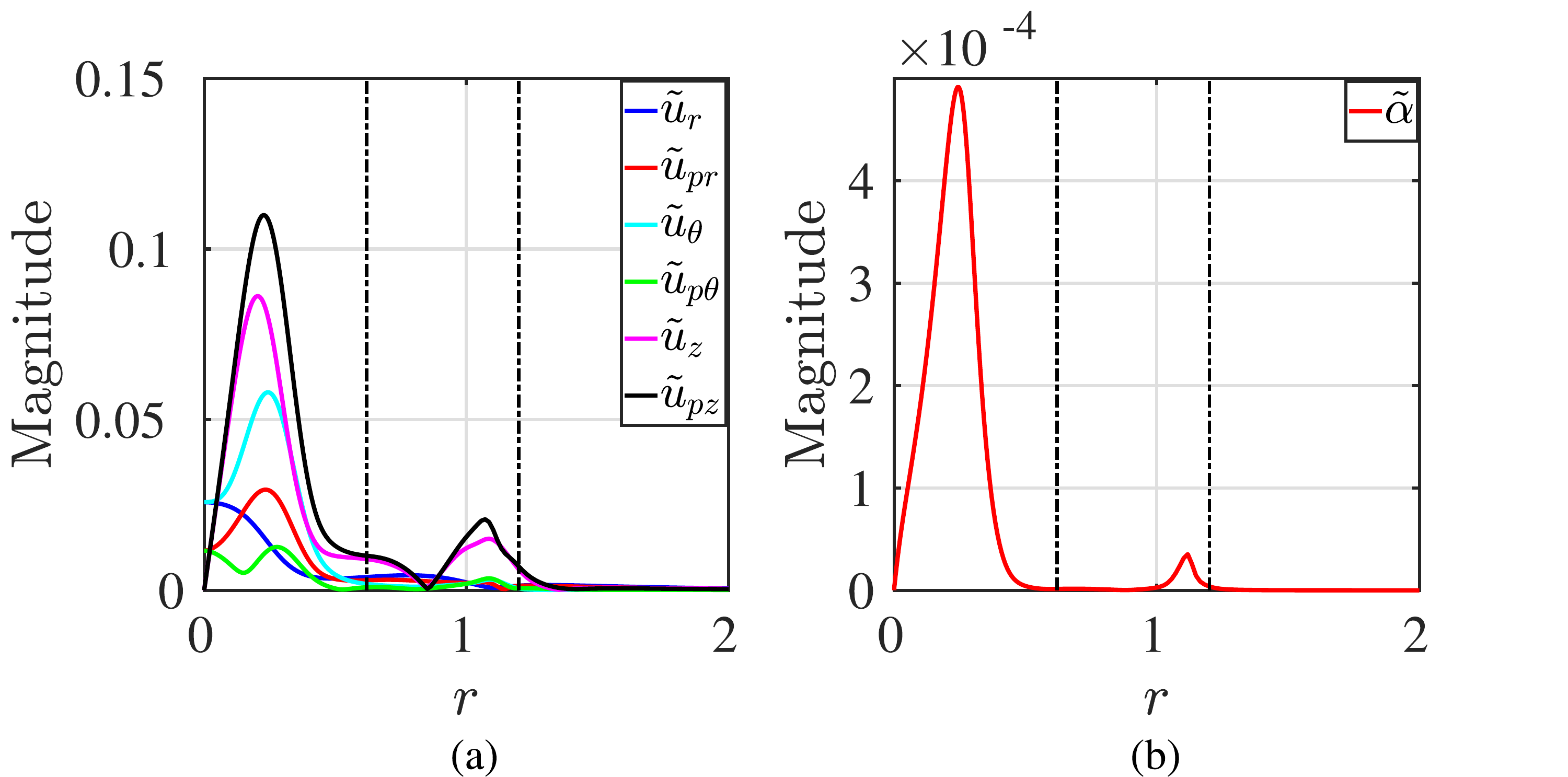}
	\caption{Eigenmode shapes of particle-laden swirling jet at Stokes number $St=1$, of the centre mode. (a) shows the radial, azimuthal, axial and (b) particle concentration eigenmode shapes . The dotted line specifies the nominal position of the inner and outer shear layers. The magnitude of the eigenmodes peaks in the vicinity of the jet centreline for both the fluid and particle.  Most of the disturbance particle concentration is within the core of the vortex core. The centre mode is evaluated at $Re=10000$, local swirl number $S=0.5$, backflow parameter $\beta=0.2$ and $k=1.2$.}
	\label{stokes_1_centre_mode_eigenvectors}  
\end{figure*}
\begin{figure*}
	\centering
    \includegraphics[width=0.75\textwidth]{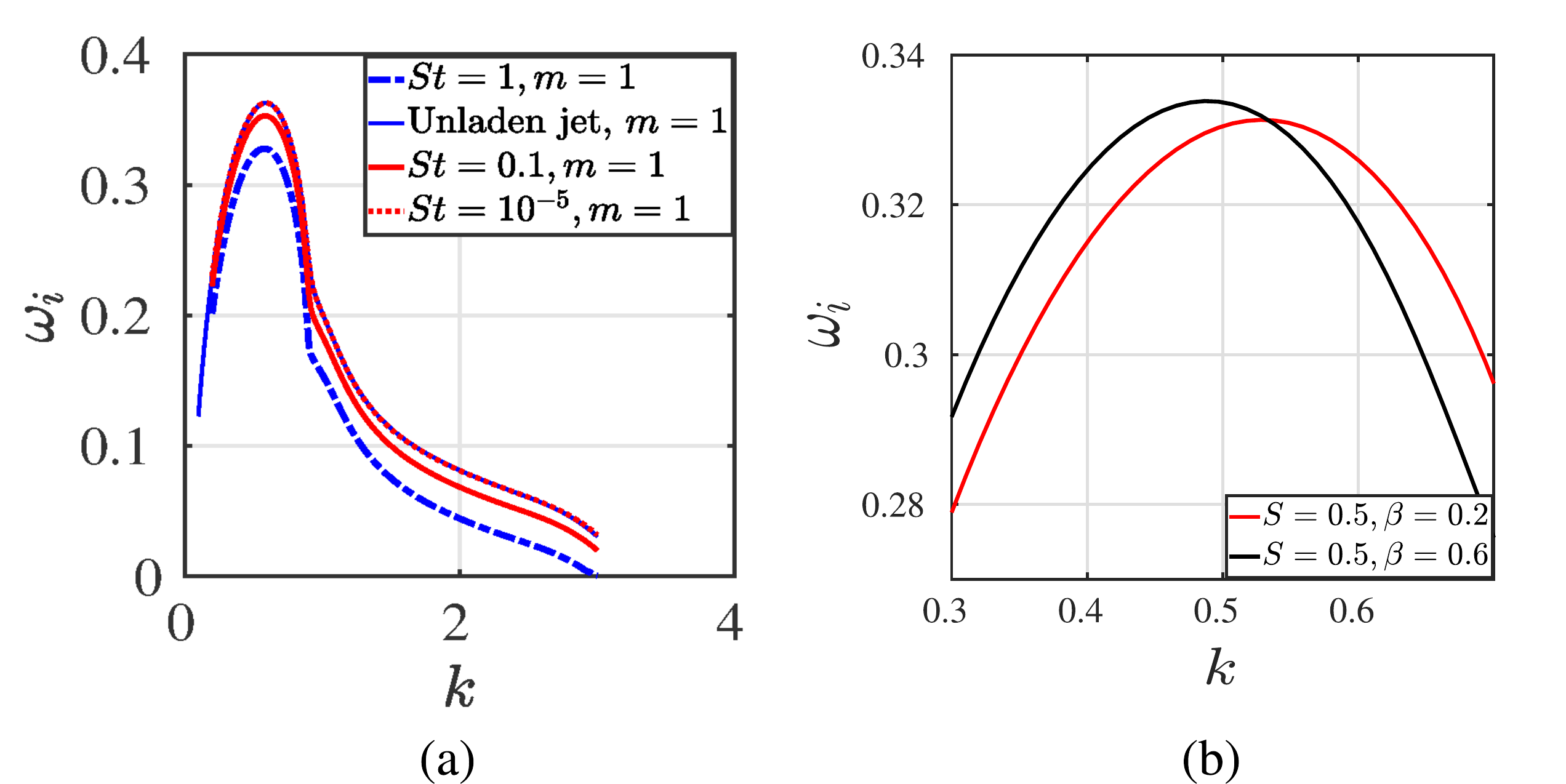} 
	\caption{(a) Growth rates of centre mode (for $m=1$ mode) of the particle-laden swirling jet in comparison to the unladen swirling jet. At low Stokes number, growth rate with particles is almost equal to the unladen jet owing to negligible drag at low Stokes numbers. As Stokes number is increased to $St=1,10$, the growth rate with particles is lower than the unladen jet. (b) Shows the variation of the growth rate of the unstable centre mode at different backflow parameters ($\beta=0.2$ and $\beta=0.6$) at a constant swirl number. Growth rate increases as the backflow parameter is increased. }
	\label{stokes_1_centre_mode_dispersion_relation}  
\end{figure*}

\begin{figure*}
	\centering
	\subfigure{\includegraphics[width=0.35\textwidth]{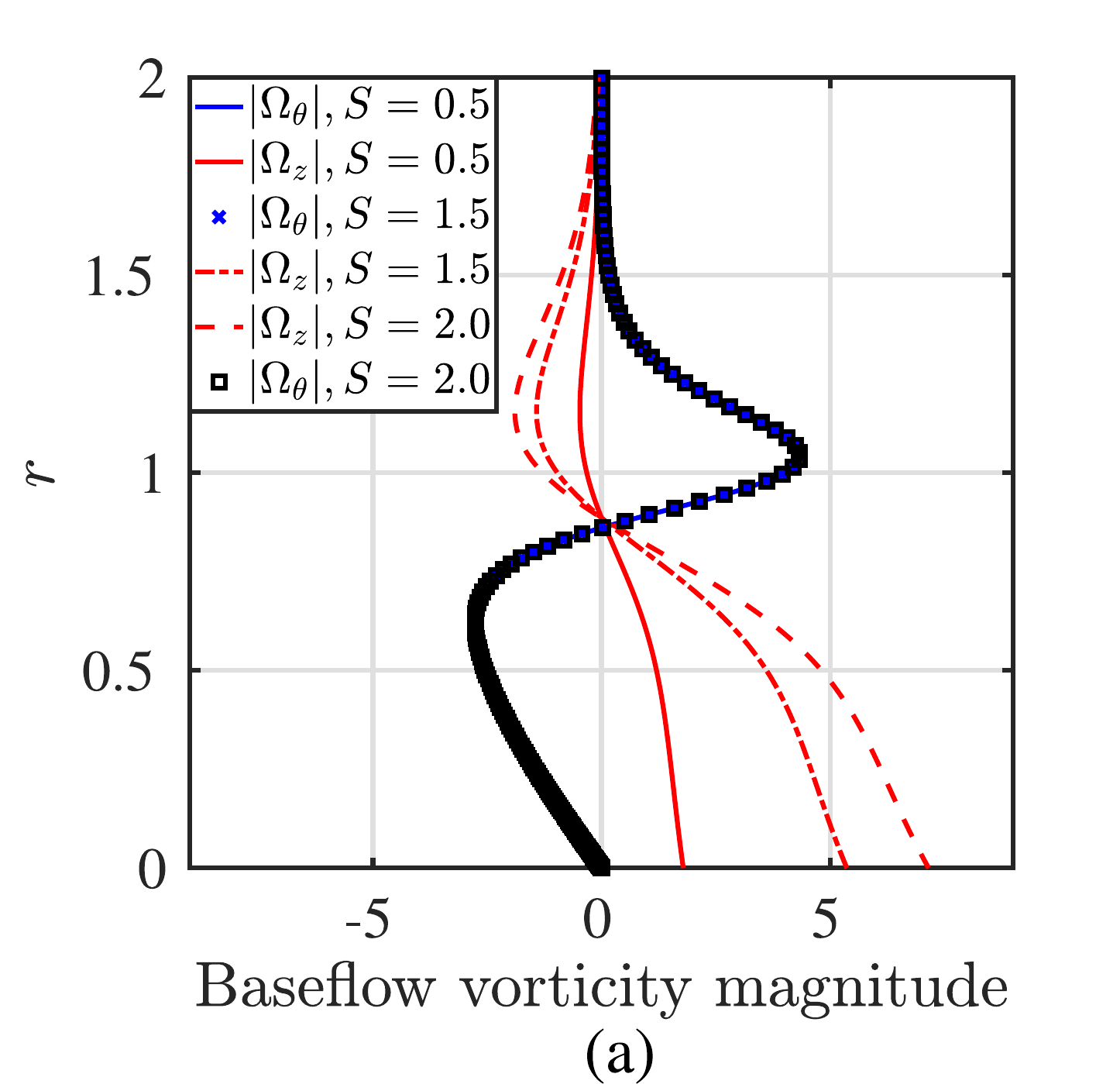}}
\subfigure{\includegraphics[width=0.35\textwidth]{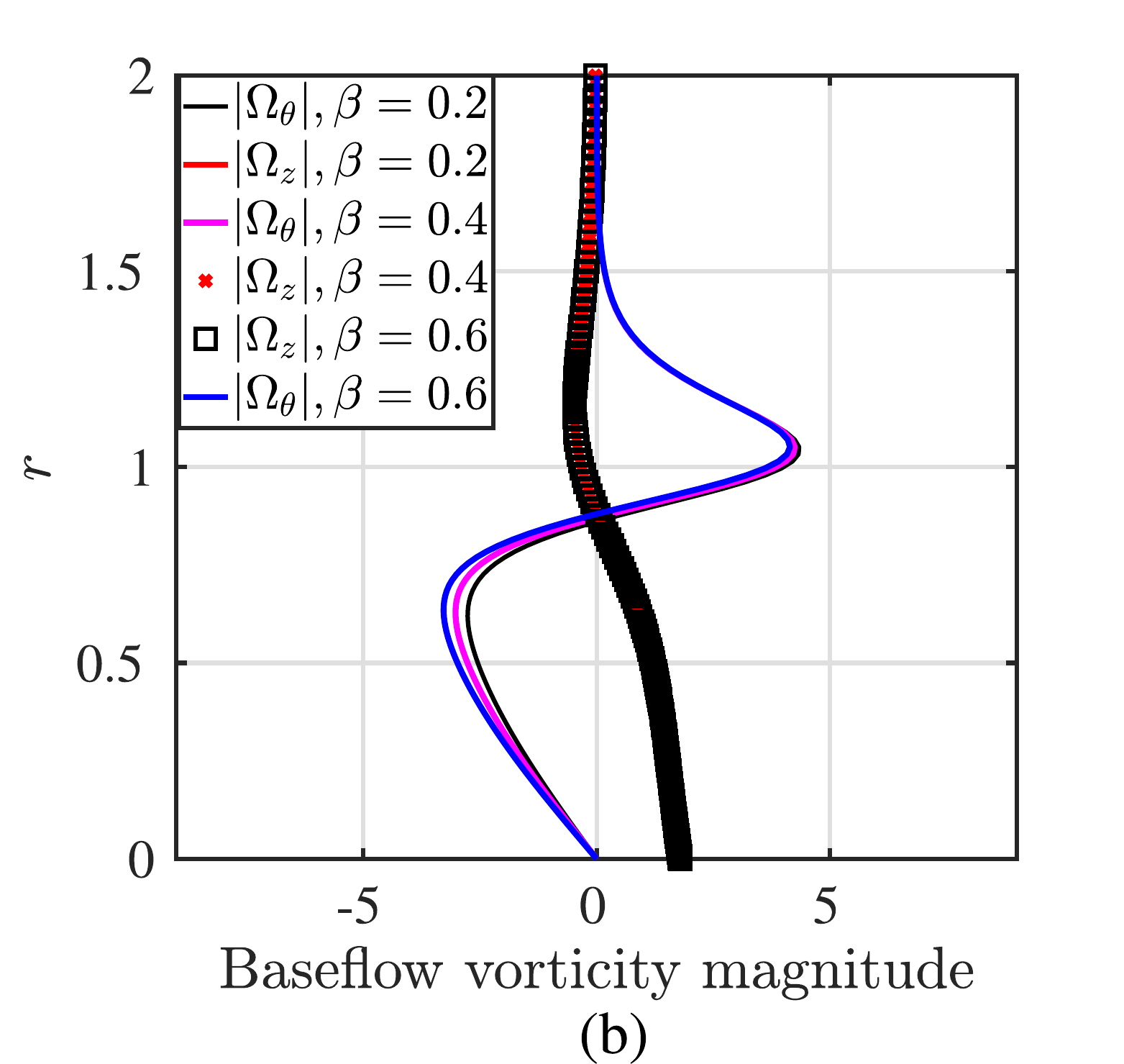}}
	\caption{Variation of axial and azimuthal vorticity for different swirl number ($S$) and the strength of the reverse flow given by
	  the backflow parameter ($\beta$). (a) varying swirl number with $\beta=0.2$. (b) varying backflow parameter $\beta$ at $S=0.5$.}
	\label{swirl_vorticity}  
\end{figure*}

\begin{figure*}
	\centering
	\includegraphics[width=0.7\textwidth]{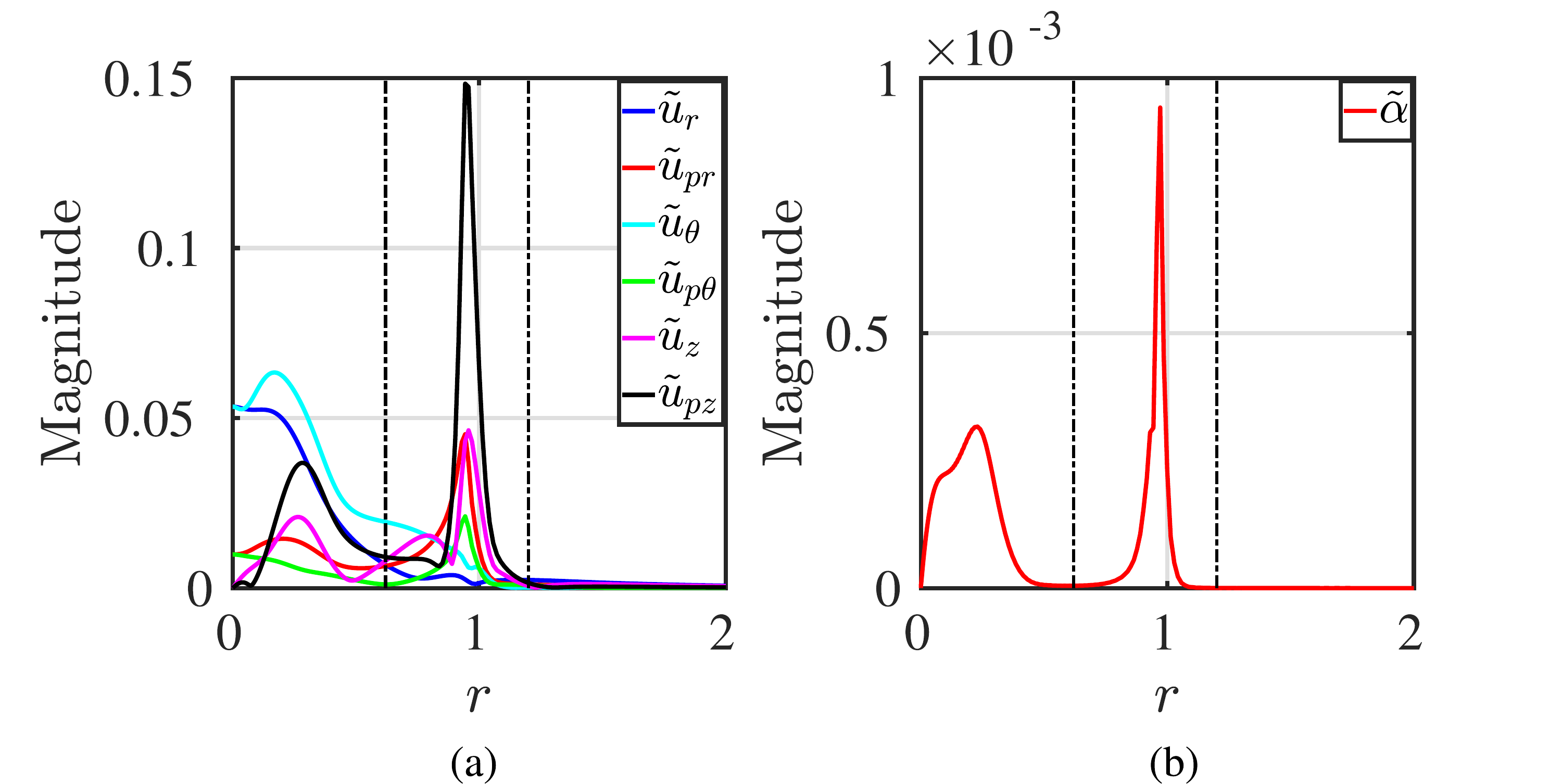} 
	\caption{ Spatial variation of the radial and azimuthal components of the fluid velocities peak in the vortex core but the corresponding particle velocities peak in the shear layer instead of the vortex core at higher swirl numbers ($S=1.5$) for the unstable centre mode. The dotted black lines indicate the nominal positions of the inner and outer shear layer. Particle concentration for the centre mode peaks in the shear layer rather than the vortex core at high swirl numbers. The axial velocity components of the fluid and particle phase peaks in the vortex core. The centre mode is evaluated at $Re=10000$, backflow parameter $\beta=0.2$.}
	\label{stokes_1_centre_mode_S_1.5}  
\end{figure*}

\begin{figure*}
	\centering
	\includegraphics[width=0.7\textwidth]{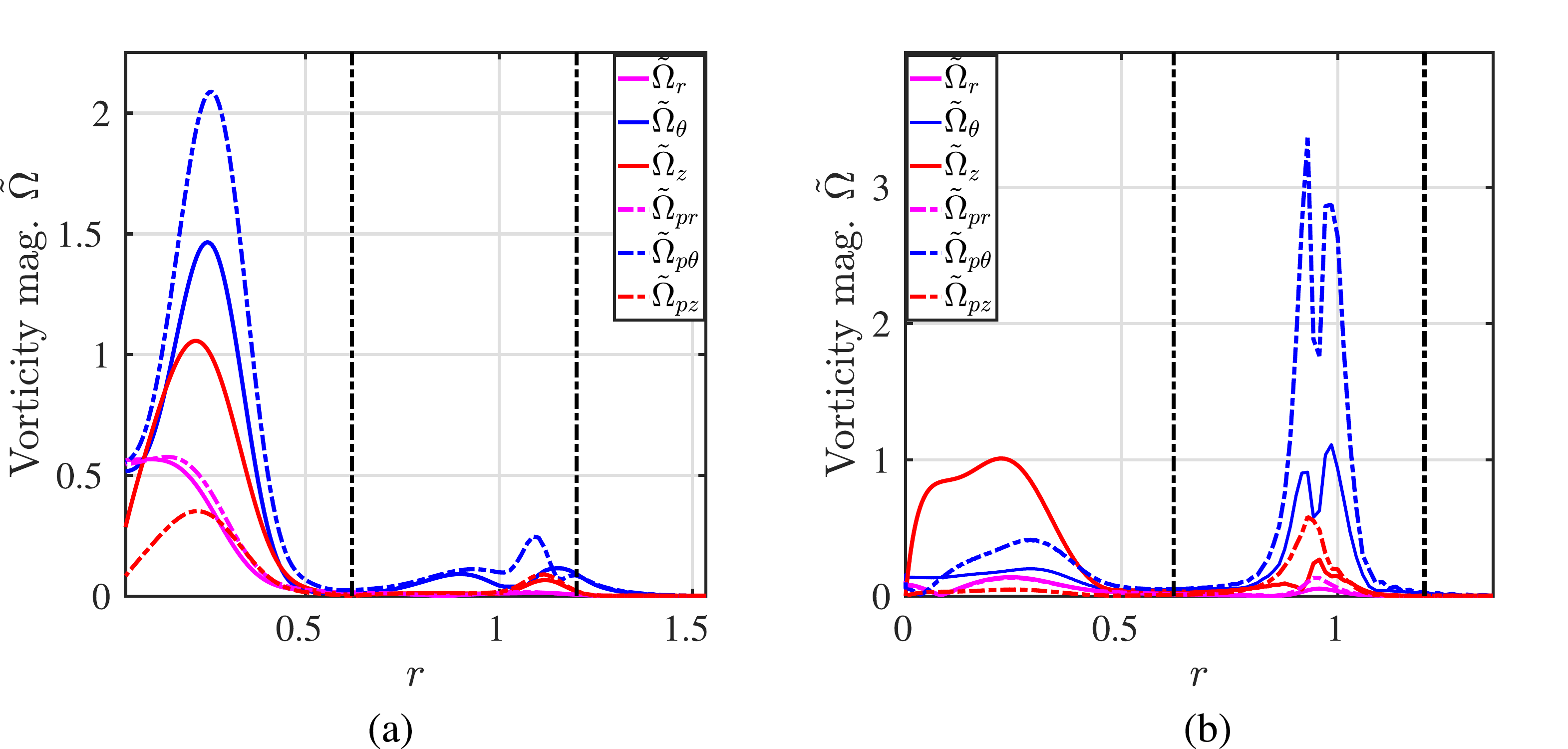} 
	\caption{ Spatial variation of the disturbance vorticity components of the fluid and particle phase for the unstable centre mode. The dotted black lines indicate the nominal positions of the inner and outer shear layer. (a) At $S=0.5$, vorticity components of the fluid and particle peaks in the vortex core. (b) At $S=1.5$, fluid  vorticity component peaks in  vortex core, while the particle vorticity components peak in the shear layer. Notice that the maximum value of $\tilde{\Omega}_{z}$ in the vortex core is almost equal to that of $\tilde{\Omega}_{\theta}$ in the shear layer, due to the fact that the base state axial vorticity is a function of swirl number, while the base state azimuthal vorticity is independent of the swirl number.}
	\label{stokes_1_centre_mode_S}  
\end{figure*}

\subsection{Centre modes}
\label{subsub:centre modes}
\begin{figure*}
	\centering
		\includegraphics[width=0.28\textwidth]{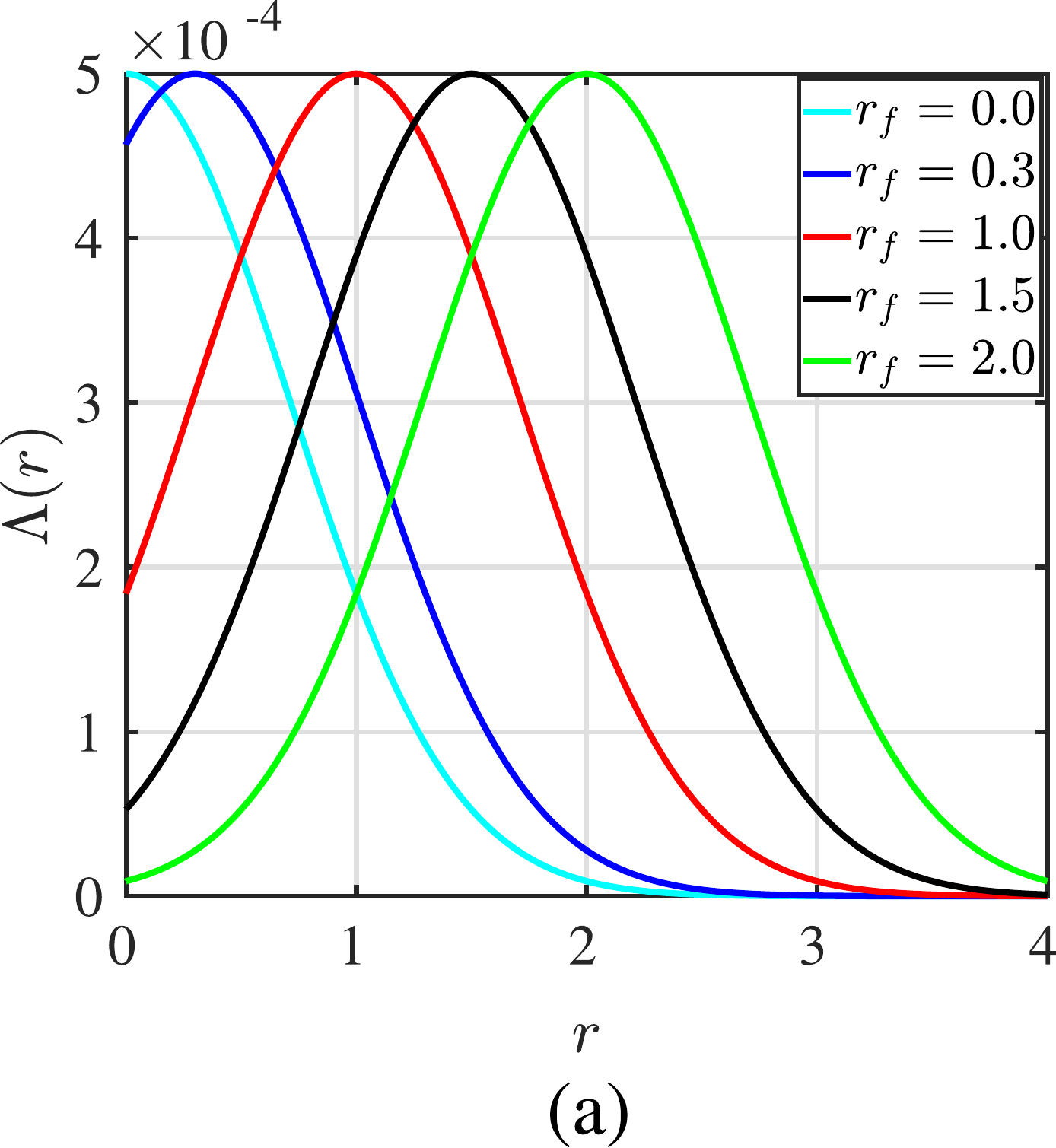}
		\includegraphics[width=0.3\textwidth]{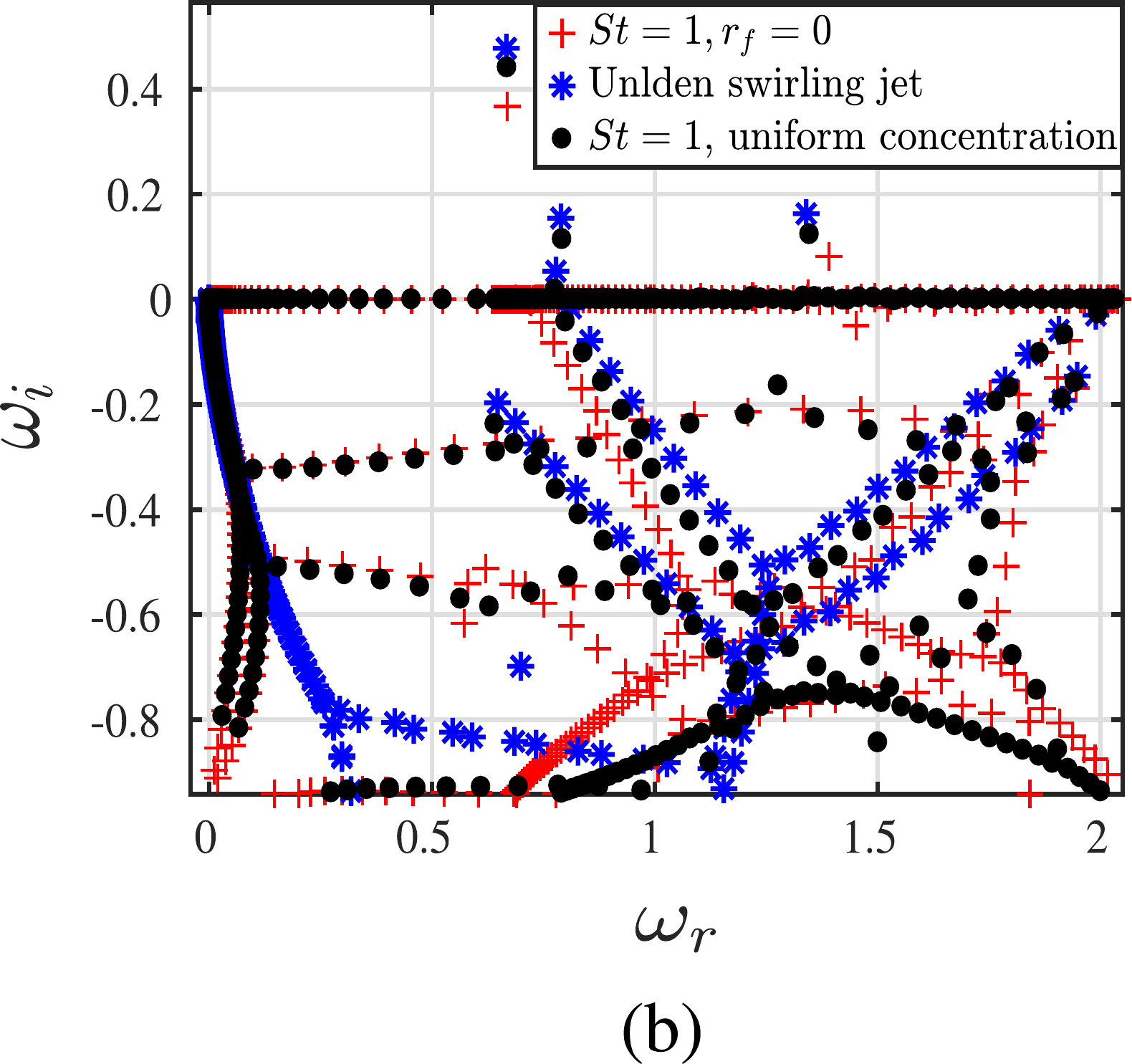}
		\includegraphics[width=0.29\textwidth]{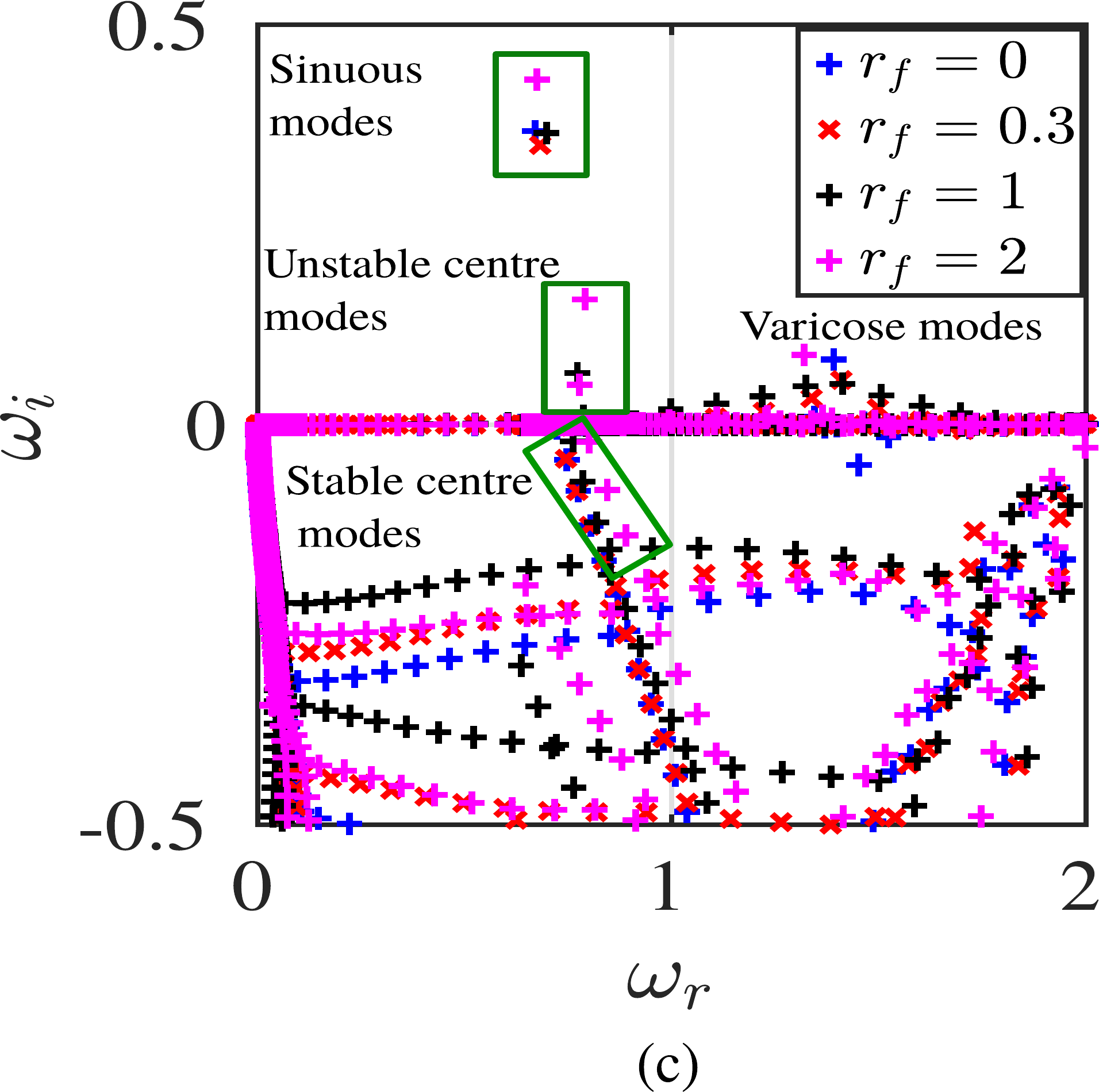}
		\caption{Baseflow particle volume concentration profiles for $r_{f} = 0, 0.3, 1.0, 1.5$ and $2$. These values represent the location of the peak of the concentration profile on the centreline, within the vortex core, in the inner shear layer, in the outer shear layer and outside the shear layer respectively. (b) Comparison of the growth rate with variable concentration with $r_{f}=0$, unladen laden swirling jet and uniform particle concentration. Centre modes are stable for $r_{f}=0$. (c) The eigenspectra for various values of $r_{f}$. Centre modes are stable if $r_{f}$ lies within the vortex core and unstable when it lies in the shear layer.}
	\label{oberliethner_VPL_CPL_unladen}  
\end{figure*}

\begin{figure*}[!ht]
	\centering
 		\includegraphics[width=0.4\textwidth]{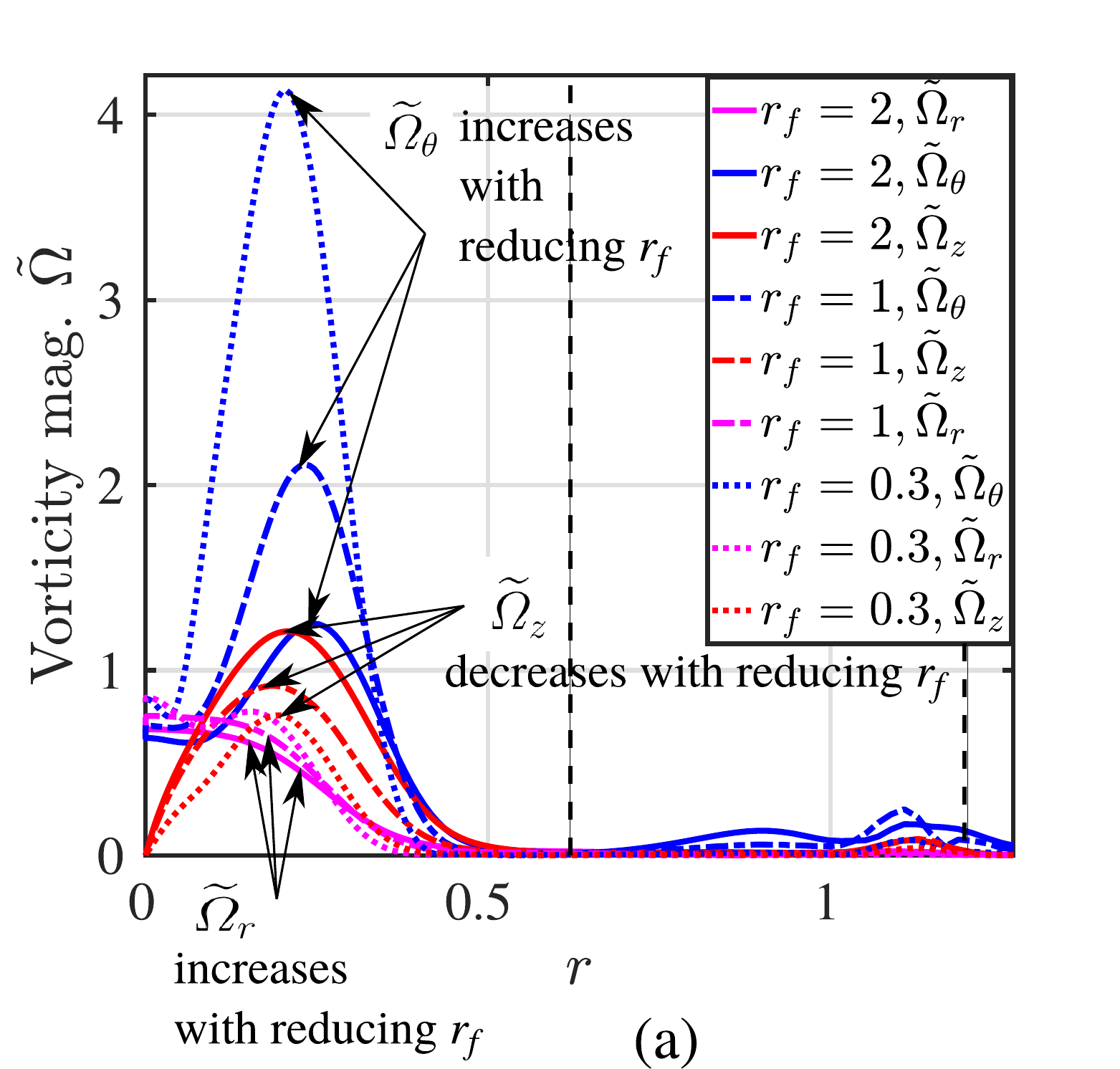}
	\includegraphics[width=0.46\textwidth]{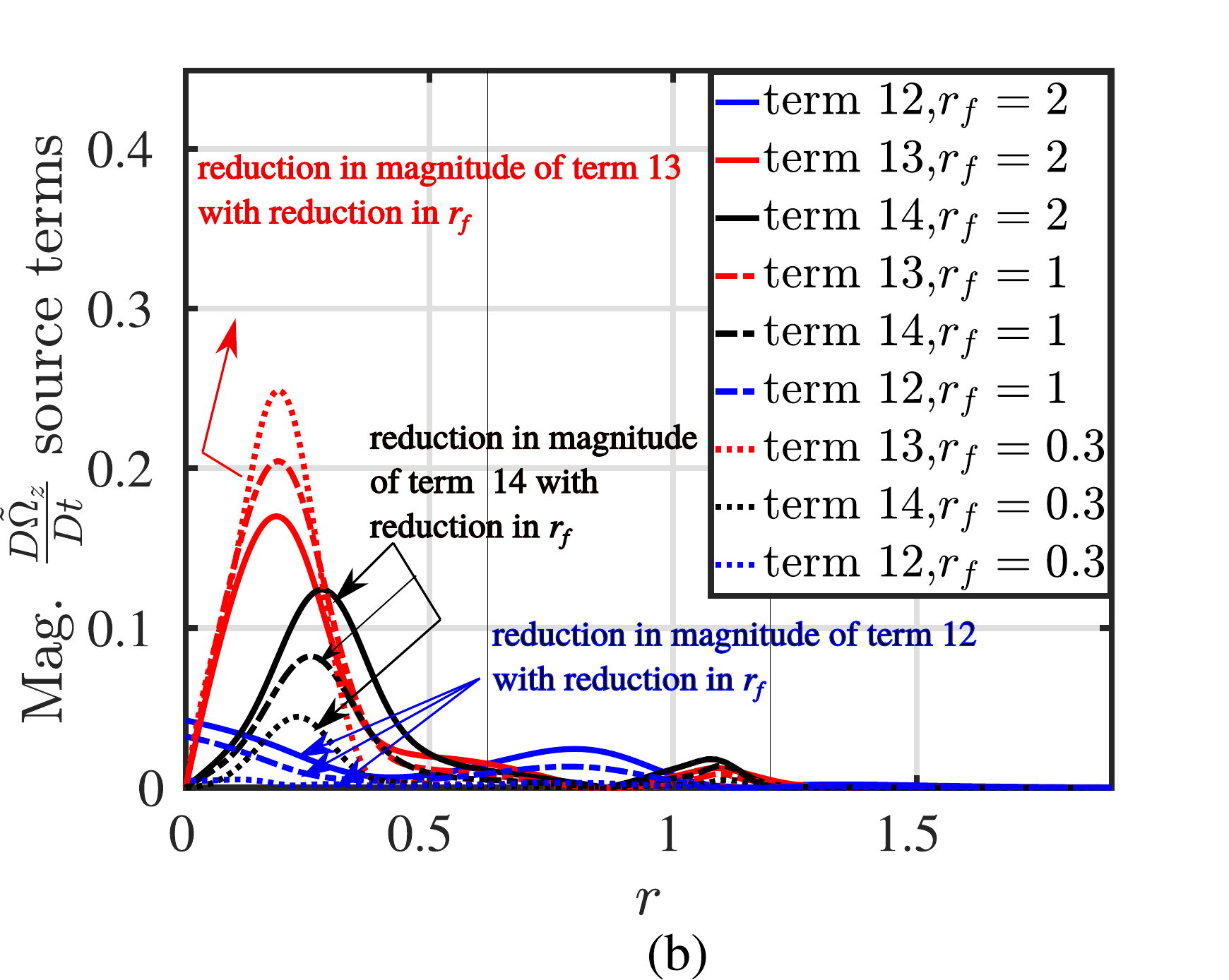}
	\caption{(a)Spatial variation of the vorticity components for the centre mode for various values of $r_{f}$. As $r_{f}$ is reduced from 2 to $0.3$, the axial vorticity reduces while the radial and azimuthal vorticity increases. (b) Spatial variation of the source terms $\protect\circled{12},\protect\circled{13}$ and $\protect\circled{14}$ in the vorticity budget equation in the axial direction Eq.(\ref{eq:axial_vorticity_budget}). Reduction in $r_{f}$ from $2$ to $0.3$ results in the decrease in the terms $\protect\circled{12},\protect\circled{13}$ and $\protect\circled{14}$. The dotted line specifies the nominal position of the inner and outer shear layers.}
	\label{vpl_vorticity_rf}  
\end{figure*}

\begin{figure*}[!ht]
	\centering
	\includegraphics[width=0.4\textwidth]{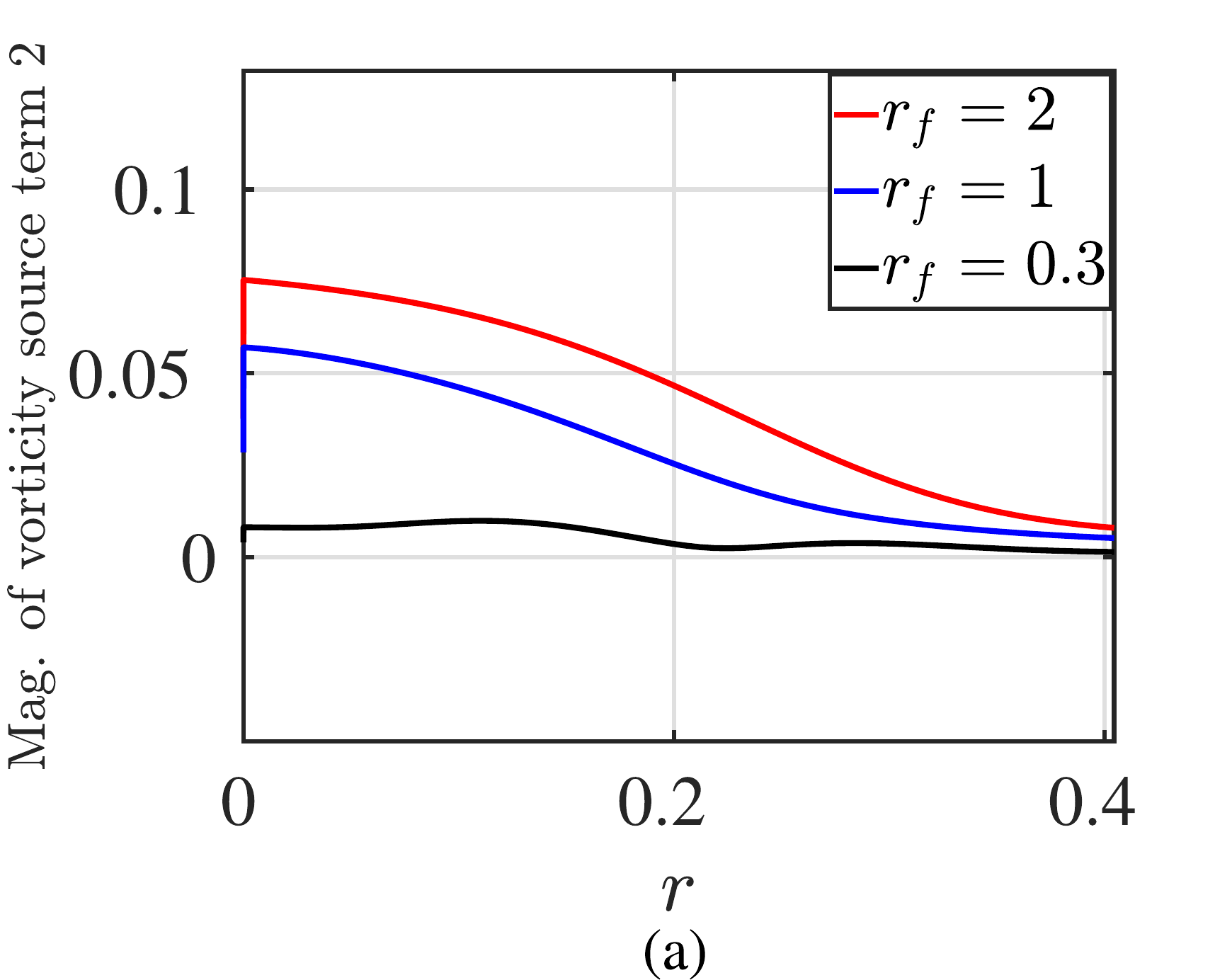}
	\includegraphics[width=0.4\textwidth]{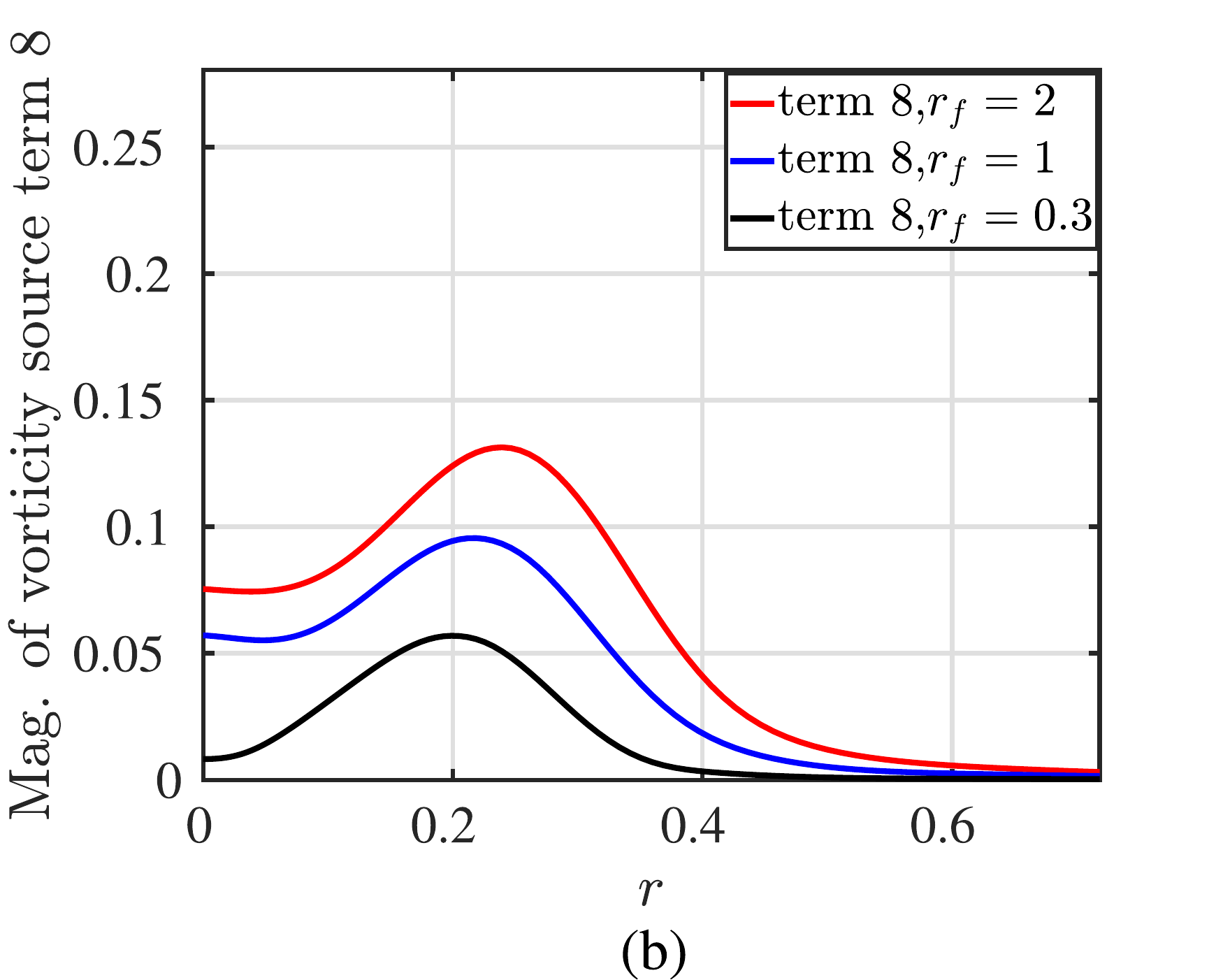}
	\caption{Spatial variation of the source terms for the centre mode (a) term $\protect\circled{2}$ and (b) term $\protect\circled{8}$ in the vorticity budget equation in the radial and azimuthal direction Eq.(\ref{eq:radial_vorticity_budget}) and Eq.(\ref{eq:azimthal_vorticity_budget}) respectively. Reduction in $r_{f}$ from $2$ to $0.3$ results in the decrease of these terms.}
	\label{term_2_and_8_vorticity_rf}  
\end{figure*}

\begin{figure*}[!ht]
	\centering
		\includegraphics[width=0.6\textwidth]{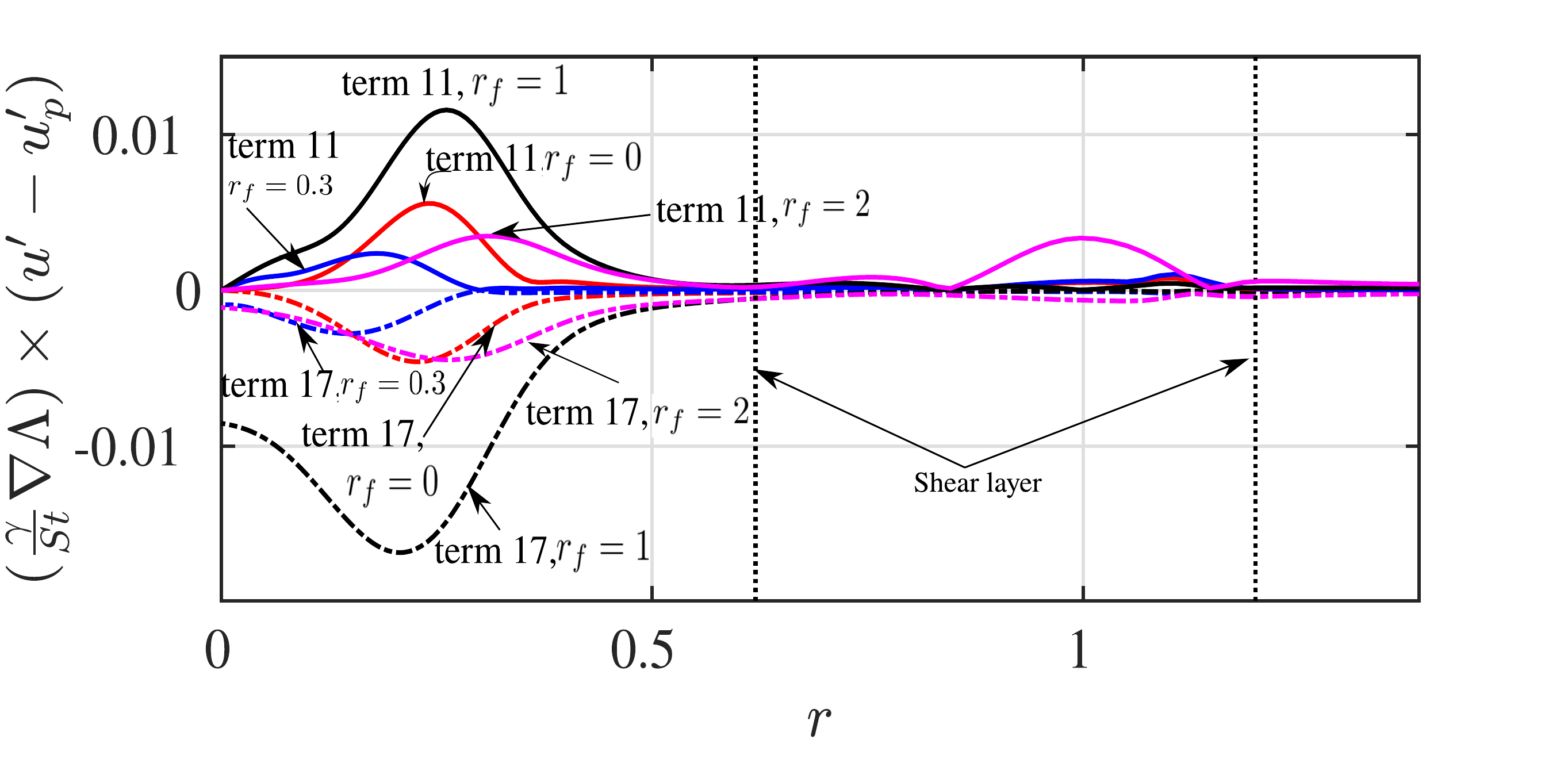}
	\caption{Spatial variation of additional terms due to the presence of gradients in the baseflow particle concentration, given by terms $\protect\circled{11}$ and $\protect\circled{17}$ in the azimuthal and axial direction respectively. For the centre mode, as $r_{f}$ reduces from $1$ to $0$, the terms $\protect\circled{11}$ and $\protect\circled{17}$ reduces. Notice that this term has no contribution from the radial component since the baseflow concentration profile is locally parallel.   }
	\label{vpl_additional_terms}  
\end{figure*}

In the preceding section, we mentioned briefly the structure of the eigenspectra for swirling jet in a swirl combustor.  
In figure(\ref{low_stokes_oberleithner_spectra}a,b), we note the presence of the centre mode for the unladen, low Stokes and intermediate Stokes number. According to Heaton's \cite{Heaton_2007} criteria (for incompressible, single phase swirling flows), the centre mode is temporally unstable, if the parameter $H_{0}<0$ in Eq.(\ref{Heaton_criteria}).
 For the present baseflow parameters, the centre mode is unstable only for $m>0$, i.e, only co-rotating modes have centre mode instability. Counter rotating centre modes  ($m<0$) are temporally stable. In the particle-laden case, we see a similar behaviour, that for counter rotating modes $m<0$, the centre modes are stable for $m=-1$ eigenspectra (not shown here). We therefore restrict the present study to $m=1$ alone. Figure(\ref{stokes_1_centre_mode_eigenvectors}) shows the velocity and particle concentration eigenfunctions. As typically seen in the case of centre modes in swirling flows (Eg. Batchelor "q" vortex stability studied by Heaton\cite{Heaton_2007}, Le Diz{\`e}s and Fabre\cite{Fabre},\cite{Fabre_TCFD}) the peak magnitude of the velocity eigenmodes occur in the vortex core (vicinity of the jet centreline) and reduces radially outward. This is made clear by indicating the positions of the inner and outer shear layers by the dotted lines for reference. Figure(\ref{stokes_1_centre_mode_eigenvectors}d) shows the radial variation of the particle concentration eigenfunctions. We note the variation in particle concentration occurs mainly in the vortex core and almost none in the shear layer. 
 For the first helical mode ($m=1$), we know from Eq.(\ref{centreline_particle}), $\tilde{\alpha}=0$,  at the jet centreline, while the particle radial and azimuthal velocity ($\tilde{u}_{r}, \tilde{u}_{pr}$) is non-zero. This indicates that particles initially on the jet centreline migrate away from the centreline towards the core for short times during which linear stability analysis is valid. 
 Figure(\ref{stokes_1_centre_mode_dispersion_relation}a) shows temporal growth rate behavior at different Stokes numbers. We see that for low Stokes numbers, the temporal growth rates are almost equal to that of the unloaded growth rate because of negligible drag force. As the Stokes number is increased to $St=1$, the Stokes drag increases and results in the reduction of growth rate. Figure(\ref{stokes_1_centre_mode_dispersion_relation}b) shows the influence of the backflow parameter ($\beta$) on the unstable centre mode. Keeping the swirl number constant ($S=0.5$), the backflow strength is increased ($\beta=0.2,0.6$) and we see that the growth rate of the centre mode increases. 
Next we study the effect of swirl number on the unstable centre mode. From figure(\ref{swirl_vorticity}) shows that the baseflow azimuthal vorticity is strongly influenced by the backflow parameter $\beta$ which enhances the axial KH.  From the expressions of the baseflow azimuthal and axial velocities, we see that the axial vorticity is influenced by the swirl number while the azimuthal vorticity is independent of the swirl number.  As mentioned previously, at moderate swirl numbers ($S=0.5$), both the fluid and particle velocities peak in the vortex core, while it decays as we move towards the shear layer. But as the swirl number is increased to $1.5$ and further, we notice that although the fluid velocities peak in the vortex core, the particle velocities peak in the shear layer. The increase in swirl number results in the magnitude of the concentration peaking in the shear layer and not in the vortex core (as in the case of moderate swirl number of $S=0.5$). This is seen in figure(\ref{stokes_1_centre_mode_S_1.5}) for swirl number of $S=1.5$. 
Figure(\ref{stokes_1_centre_mode_S}) shows the fluid and particle vorticity generation rate in the radial, azimuthal and axial directions for different swirl numbers. We see that at moderate swirl numbers, the vorticity components of the fluid and particle peaks in the vortex core. The fluid vorticity component peaks in vortex core, while the particle vorticity components peak in the shear layer. Notice that the maximum value of $\Omega_{z}$ in the vortex core is almost equal to the maximum value of $\Omega_{\theta}$  in the shear layer, due to the fact that the base-state axial vorticity is a function of swirl number, while the base-state azimuthal vorticity is independent of the swirl number.
We investigate the effect of baseflow variable particle loading on the temporal growth rate of the centre mode. Consider baseflow concentration profile given by,
\begin{equation}
	\Lambda\left(r\right)=\overline{\alpha}\exp\left(-\frac{\left(r-r_{f}\right)^{2}}{l_{p}}\right).
	\label{VPL_basestate}
\end{equation}
where $\overline{\alpha}$ is a constant $\left(10^{-4}\right)$, $r_{f}$ is the offset of the peak of the concentration profile from the jet centreline, $l_{p}$ is the steepness parameter $\left(0.2 \mbox{ in the present case}\right)$. 
 Figure(\ref{oberliethner_VPL_CPL_unladen}) shows the comparison of the eigenspectra for the variable concentration  profile with $r_{f}=0$, uniform concentration $\Lambda=10^{-4}$ and the unladen swirling jet.  We see that for $r_{f}=0$ (peak of the concentration profile is on the centreline)  and $r_{f}=0.3$ (peak lies in the CRZ) that, the centre modes are stable and only the shear layer modes (sinuous and varicose modes) are unstable. While for $r_{f}=1$ (inside the shear layer) and $r_{f}=2$ (outer region of the jet), the centre modes and shear layer modes remain unstable. As the offset distance is increased, the peak of the concentration profile is moved away from the centreline, the growth rates increases. In other words, as the peak of the baseflow concentration profile is shifted from the outer shear layer to the vortex core, the centre mode becomes less unstable and become stable when $r_{f}$ lies inside the vortex core and on the centreline as well. 
 Next, we discuss the effect of particle addition on the linearized vorticity transport. Consider the spatial variation of the disturbance vorticity components as shown in figure(\ref{vpl_vorticity_rf}a). We see that as we shift the peak of the baseflow concentration profile from the inner shear layer to the vortex core of the jet, the magnitude of the axial vorticity decreases while the magnitude of the azimuthal and radial vorticity increases. Consider the vorticity budget equation in the axial direction i.e., Eq.(\ref{eq:axial_vorticity_budget}).  The left hand side represents the net change of axial perturbation vorticity given by $\frac{D\Omega_{z}^{'}}{Dt}$. Term $\circled{12}$ in Eq.(\ref{eq:axial_vorticity_budget}) can be written as,   
\begin{equation}
-\left(\frac{d^{2}U_{\theta}}{dr^{2}}+\frac{1}{r}\frac{dU_{\theta}}{dr}-\frac{U_{\theta}}{r^{2}}\right)\tilde{u}_{r}= -{u}_{r}'\frac{d\Omega_{z}}{dr}.
\end{equation}
Term $\circled{12}$ is basically the advection of the base state axial vorticity gradient by the perturbation radial velocity. Similarly term $\circled{13}$ can be written as,
\begin{equation}
ik\left(\frac{dU_{\theta}}{dr}+\frac{U_{\theta}}{r}\right)\tilde{u}_{z}= \Omega_{z}\frac{\partial u_{z}'}{\partial z}.
\end{equation} 

which is the perturbed vortex stretching term. 
As the peak of the particle concentration is shifted towards the vortex core, for instance when $r_{f}=0.3$, we see reduction in the magnitude of terms $\circled{12}$ and $\circled{13}$ as shown in figure(\ref{vpl_vorticity_rf}b). 
Similarly term $\circled{14} = -ik\tilde{u}_{\theta}\frac{dU_{z}}{dr} = \Omega_{\theta}\frac{\partial u_{\theta}'}{\partial z}$ is the perturbed vortex stretching term that sees a reduction in magnitude as $r_{f}$ reduces from $2$ to $0$ as shown in figure(\ref{vpl_vorticity_rf}b), a decrease in the net generation of perturbation vorticity in the axial direction is noted. 
Next consider the vorticity budget equations in the radial and azimuthal directions given by Eqs.(\ref{eq:radial_vorticity_budget},\ref{eq:azimthal_vorticity_budget}) respectively. Consider the term $\circled{2}$ in Eq.(\ref{eq:radial_vorticity_budget}) which can be written as,

\begin{equation}
\left(\frac{dU_{\theta}}{dr}+\frac{U_{\theta}}{r}\right)ik\tilde{u}_{r} = \Omega_{z}\frac{\partial u_{r}'}{\partial z}.
\end{equation}

which is the radial component of the perturbed vortex stretching term. 
We see from figure(\ref{term_2_and_8_vorticity_rf}a) that reduction in $r_{f}$ from $2$ to $0.3$ lowers the contributions of term $\circled{2}$. Consider the vorticity budget equation in the azimuthal direction given by Eq.(\ref{eq:azimthal_vorticity_budget}). 
Figure(\ref{term_2_and_8_vorticity_rf}b) shows that term $\circled{8}$ decreases in magnitude as $r_{f}$ is reduced from $2$ to $0.3$. The decrease in the axial perturbation vorticity is accompanied by an increase in the azimuthal perturbation vorticity generation as seen in figure(\ref{vpl_vorticity_rf}a). Furthermore, due to the presence of non-uniformity in the baseflow concentration, terms $\circled{11}$ and $\circled{17}$ modify the perturbation vorticity components in the azimuthal and axial direction respectively. This is shown in figure(\ref{vpl_additional_terms}). Note that the radial component of this term is zero since the cross product of the radial component of the base-state particle concentration with the perturbation slip velocity vector is zero because the concentration profile is locally parallel.


\section{Conclusions}
\label{sec:conclusion}
We explore the linear local stability characteristics of swirling jet laden with particles in an annular swirl combustor. Volume averaged equations in cylindrical coordinates in the dilute suspension regime are considered here.  The closure term arising from volume averaging is neglected assuming the suspension to be dilute and particle Reynolds number to be very small. Prior local analysis by Manoharan \cite{kiran_thesis_2019} on unladen swirling jet revealed the presence of three unstable modes, namely, centre modes(mainly driven by centrifugal instability), sinuous and varicose modes which are shear layer modes driven mainly by KH instability mechanism.  Eigenspectra of the particle-laden jet with uniform particle concentration shows three unstable modes (centre, sinuous and varicose modes) and a new set of stable modes which are absent in the unladen jet eigenspectrum. As Stokes number is increased, the growth rates of the centre and shear layer modes  reduces compared to that of the unladen swirling jet. Magnitude of the velocity eigenmodes peaks in the vortex core and reduces radially outward. Variation in the particle concentration occurs mostly in the vortex core and almost none in the shear layer.  But as the swirl number is increased to $1.5$ and further, it is seen that although the fluid velocities peak in the vortex core, the particle velocities peak in the shear layer. The increase in swirl number results in the particle concentration magnitude peaking in the shear layer and not in the vortex core (as in the case of moderate swirl number of $S=0.5$). The effect of increase in backflow parameter is to increase the growth rate of the centre mode.

	We look at the effect of non-uniformity of the baseflow particle concentration on the stability of the swirling jet. The fluid and particle base flow velocities are assumed to be equal, which implies drag force is zero for the base state. This allows us to impose any locally parallel baseflow concentration profile. We consider a Gaussian concentration profile with its location of the maximum varied, viz, located on the centreline, in the vortex core, in the shear layer and outside the shear layer.  It is found that when the peak of the concentration profile lies within the vortex core, the growth rate of the centre modes reduces. As the peaks are shifted further away from the core (i.e., within the shear layers) and outside the shear layer, the growth rate of centre mode increases but is still smaller than the unladen flow. Sinuous and varicose modes remain unstable regardless of where the peak is located, although the growth rates are smaller than the unladen flow. Ring modes remain stable as in the case of unladen flow and uniform concentration. A detailed vorticity budget analysis reveals that as the location of the peak is brought closer to the jet centreline (inside the vortex core), there is a decrease in the net generation of perturbation vorticity in the axial direction. This is accompanied by an increase in the generation of radial and azimuthal perturbation vorticity. The radial, azimuthal velocities and particle concentration fields for the present calculation from the local profile with particles at $St=1$, suggest that the effect of the addition of particles is to reduce the temporal growth rate of the disturbances. A detailed investigation along the lines of a fully global stability analysis (on a three dimensional baseflow) has to be carried out in order to get a complete picture of the instability mechanism and to understand the effect of particles on the stability of the swirling jet. We aim to perform these calculations in the future. Temporal analysis performed here should be viewed bearing in mind that these calculations are valid in the near field of the nozzle, as the baseflow profiles further downstream are found to be non-parallel (Manoharan \cite{kiran_etal_2020}). Although the assumption of the particle baseflow velocity being equal to the baseflow fluid velocity is appropriate at low Stokes numbers, for $St=1$,  experimental investigation supplemented with numerical simulations must be performed to see if the particles undergo preferential concentration, where the particles migrate towards regions of high strain rate and low vorticity. In such cases, the particle baseflow field has to be modeled accordingly. However, the results of our calculations are expected to help guide the numerical simulation by providing better initial conditions and to interpret the early stages of the evolution of the particle-laden swirling jet with the configuration considered by Manoharan\cite{kiran_thesis_2019}. From an experimental point of view as well, the present linear stability calculations open up new avenues to conduct studies on this configuration so that a more accurate description of the effect of particles on the swirling jet can be obtained. 
\appendix
\section{Matrix form for variable concentration particle laden swirling jet}
\label{appendix_a}
The linearized equations are arranged in a row wise fashion as continuity equation, $r$ momentum equations in the radial, azimuthal and axial directions for the continuous phase and momentum equations along with the continuity equation for the dispersed phase. The elements of $A$ and $B$ matrices are given below,
\[
A=\begin{bmatrix}A_{11} & A_{12} & A_{13} & 0 & 0 & 0 & 0 & A_{18}\\
A_{21} & A_{22} & 0 & A_{24} & A_{25} & 0 & 0 & A_{28}\\
A_{31} & A_{32} & A_{33} & A_{34} & 0 & A_{36} & 0 & A_{38}\\
A_{41} & 0 & A_{43} & A_{44} & 0 & 0 & A_{47} & A_{48}\\
A_{51} & 0 & 0 & 0 & A_{55} & A_{56} & 0 & 0\\
0 & A_{62} & 0 & 0 & A_{65} & A_{66} & 0 & 0\\
0 & 0 & A_{73} & 0 & A_{75} & 0 & A_{77} & 0\\
0 & 0 & 0 & 0 & A_{85} & A_{86} & A_{87} & A_{88}
\end{bmatrix}
\]

\[
B=\begin{bmatrix}0 & 0 & 0 & 0 & 0 & 0 & 0 & {\color{black}-i}\\
i\left(1{\color{black}-\Lambda}\right) & 0 & 0 & 0 & 0 & 0 & 0 & 0\\
0 & i\left(1{\color{black}-\Lambda}\right) & 0 & 0 & 0 & 0 & 0 & 0\\
0 & 0 & i\left(1{\color{black}-\Lambda}\right) & 0 & 0 & 0 & 0 & 0\\
0 & 0 & 0 & 0 & i & 0 & 0 & 0\\
0 & 0 & 0 & 0 & 0 & i & 0 & 0\\
0 & 0 & 0 & 0 & 0 & 0 & i & 0\\
0 & 0 & 0 & 0 & 0 & 0 & 0 & i
\end{bmatrix}
\]
\begin{equation*}
\begin{aligned}
& A_{11}=\left(1-\Lambda\right)\frac{d}{dr}{-\frac{d\Lambda}{dr}}+\frac{1-\Lambda}{r}, ~ A_{12}=\left(\frac{1-\Lambda}{r}\right)im, ~ A_{13}=ik\left(1-\Lambda \right), ~ A_{18}=\left(-ikU_{z}-\frac{imU_{\theta}}{r}\right).&\\
&A_{21}\mbox{ can be split into three parts as } A_{21}=\left(A_{211}+\frac{1}{Re}A_{212}+A_{213}\right) \mbox{ where,} &\\
&A_{211}=\left(1-\Lambda\right)ikU_{z}+\left(\frac{1-\Lambda}{r}\right)imU_{\theta}, ~A_{213}=-\frac{\Lambda\gamma}{St}, \mbox{ and } &\\
&A_{212}=\left(1-\Lambda\right)^{2}\left(\frac{d^{2}}{dr^{2}}-\left(\frac{m^{2}}{r^{2}}+k^{2}\right)+\frac{1}{r}\frac{d}{dr}\right) -2\left(1-\Lambda\right)\frac{d\Lambda}{dr}\frac{d}{dr}-\left(1-\Lambda\right)\frac{d^{2}\Lambda}{dr^{2}}-\frac{1-\Lambda}{r}\frac{d\Lambda}{dr}-\frac{\left(1-\Lambda\right)^{2}}{r^{2}}  &\\
& A_{22}=-\frac{2\left(1-\Lambda\right)}{r}U_{\theta}+\frac{1}{Re}\left(\frac{2im}{r^{2}}\left(1-\Lambda\right)^{2}\right),  ~ A_{24}=\left(1-\Lambda\right)\frac{d}{dr}, ~A_{25}=-\frac{\Lambda\gamma}{St},   &\\
&A_{28}=-\frac{2}{Re}\left(\frac{1}{r^{2}}\left(1-\Lambda\right)imU_{\theta}\right), ~A_{32}=A_{21}, ~ A_{31}=\left(1-\Lambda\right)\frac{dU_{\theta}}{dr}+\frac{\left(1-\Lambda\right)}{r}U_{\theta}-\frac{1}{Re}\left(\frac{2im\left(1-\Lambda\right)^{2}}{r^{2}}\right).	&\\
& ~ A_{33}=ik\left(1-\Lambda\right)U_{\theta}, ~ A_{34}=\frac{i\left(1-\Lambda\right)m}{r}, ~ A_{36}=-\frac{\Lambda\gamma}{St}, ~A_{41}=\left(1-\Lambda\right)\frac{dU_{z}}{dr}, ~ A_{44}=ik\left(1-\Lambda\right), ~ A_{47}=-\frac{\Lambda\gamma}{St}. &\\
&A_{43}\mbox{ can be split into three parts as } A_{43}=\left(A_{431}+\frac{1}{Re}A_{432}+A_{433}\right) \mbox{ where,} &\\
&A_{431}= A_{211}, ~ A_{433}= A_{213}  \mbox{ and } &\\
&A_{432}=\left(1-\Lambda\right)^{2}\left(\frac{d^{2}}{dr^{2}}-\left(\frac{m^{2}}{r^{2}}+k^{2}\right)+\frac{1}{r}\frac{d}{dr}\right) -2\left(1-\Lambda\right)\frac{d\Lambda}{dr}\frac{d}{dr}-\left(1-\Lambda\right)\frac{d^{2}\Lambda}{dr^{2}}-\frac{1-\Lambda}{r}\frac{d\Lambda}{dr}  &\\
&A_{48}=\left(1-\Lambda\right)U_{z}\left(\frac{d^{2}}{dr^{2}}-\left(\frac{m^{2}}{r^{2}}+k^{2}\right)+\frac{1}{r}\frac{d}{dr}\right)-U_{z}\frac{d^{2}\Lambda}{dr^{2}}-U_{z}\frac{d\Lambda}{dr}+2\left(1-\Lambda\right)\frac{d^{2}U_{z}}{dr^{2}}+\frac{2\left(1-\Lambda\right)}{r}\frac{dU_{z}}{dr}+ &\\
&2\frac{d\Lambda}{dr}\frac{dU_{z}}{dr}+2\left(1-\Lambda\right)\frac{dU_{z}}{dr}\frac{d}{dr}   & \\
&A_{51}=-\frac{1}{St}, ~ A_{55}=\left(\frac{imU_{\theta}}{r}+iU_{z}k+\frac{1}{St}\right), ~ A_{56}=-\frac{2U_{\theta}}{r}.		& \\
&A_{62}=-\frac{1}{St}, ~ A_{65}=\left(\frac{dU_{\theta}}{dr}+\frac{U_{\theta}}{r}\right), ~ A_{66}=\left(ikU_{z}+\frac{imU_{\theta}}{r}+\frac{1}{St}\right).		&\\
&A_{73}=-\frac{1}{St}, ~ A_{75}=\frac{dU_{z}}{dr}, ~ A_{77}=\left(ikU_{z}+\frac{imU_{\theta}}{r}+\frac{1}{St}\right).		&\\
&A_{85}=\left(\frac{d\Lambda}{dr}+\Lambda\frac{d}{dr}+\frac{\Lambda}{r}\right), ~ A_{86}=\frac{\Lambda im}{r}, ~A_{87}=\Lambda ik, ~ A_{88}=\left(ikU_{z}+\frac{imU_{\theta}}{r}\right).			
\end{aligned}
\end{equation*}

\bibliographystyle{unsrt}  
\bibliography{references} 
\end{document}